\documentclass{aastex631}
\usepackage{subfigure}
\usepackage{amsmath}
\usepackage{hyperref}
\usepackage{longtable}
\usepackage{orcidlink}
\usepackage[cp1252]{inputenc}

\def\arcmin{{$^{\prime}$}}
\def\arcsec{{$^{\prime\prime}$}}
\def\ptsec{$''\mskip-7.6mu.\,$}

\def\Msun{\,{\rm M$_{\odot}$}}
\def\Lsun{\,{\rm L$_{\odot}$}}

\newcommand{\kms}{\mbox{km\,s$^{-1}$}}

\def\degr{$^{\circ}$}

\shorttitle{MIR Observations of MWC\,297}
\shortauthors{Vacca et al.}

\begin{document}

\title{Multi-wavelength Observations of MWC\,297:  Constraints on Disk Inclination and Mass Outflow}

\author{William D.\ Vacca}
\affil{SOFIA-USRA, NASA Ames Research Center, MS 232-12, Building N232, Rm.
147, Moffett Field, CA 94035-0001,
USA 
\orcidlink{0000-0002-9123-0068}}
\email{wvacca@sofia.usra.edu}

\author{G\"oran Sandell}
\affil{ Institute for Astronomy, University of Hawai`i at Manoa,  Hilo,  640 N. Aohoku Place, Hilo, HI 96720, USA
 \orcidlink{0000-0003-0121-8194}}

\begin{abstract}
MWC\,297 is a young, early-type star driving an ionized outflow and surrounded
by warm, entrained dust. Previous analyses of near- and mid-IR interferometric
images suggest that the emission at these wavelengths arises from a compact
accretion disk with a moderate ($i <$ 40\degr) inclination. 
We have obtained 5-40 $\mu$m images
of MWC 297 with FORCAST on SOFIA, as well as near-infrared spectra acquired with
SpeX on the IRTF and radio data obtained with the VLA and BIMA, and supplemented these with archival
data from Herschel/PACS and SPIRE. The FORCAST
images, combined with the VLA data, indicate that the
outflow lobes are aligned nearly north-south and are well separated. Simple
geometrical modeling of the FORCAST images suggests that the disk driving the outflow has an
inclination of $55\pm 5$\degr, in disagreement with the results of the interferometric
analyses. Analysis of the SpeX data, with a wind model, suggests the the mass
loss rate is on the order of $6.0 \pm ^{3.7}_{1.7} \times 10^{-7} M_\odot ~\mathrm{yr}^{-1}$ and the
extinction to the source is $A_V \sim 8.1 \pm^{2.5}_{1.5}$ 
mag. We have combined our data with values from the literature
to generate the spectral energy distribution of the source from $0.35~ \mu$m to $6$ cm and estimate the total luminosity. 
We find the total luminosity to be about $7900 ~ L_\odot$, if we include emission from an extended region around the star,
only slightly below that expected for a B1.5V star.
The reddening must be produced by dust along the line of sight, but distant from the star.

\end{abstract}


\section{Introduction}
As the more massive (2-10 $M_\odot$; \citealt{Waters98}) counterparts of the low-mass Classical T Tauri Stars, Herbig AeBe (HAeBe) stars 
represent a bridge between low mass young stellar objects, whose formation processes are (at least qualitatively) well characterized and understood,
and high mass stars, whose formation processes are (quite literally) still shrouded in dust and uncertainty. Therefore they represent a means of testing
whether the low mass star formation paradigm can be extended to higher mass stars. While it is clear that these young, pre-main 
sequence stars are actively accreting material from their circumstellar environments, the geometry of the material and the mechanisms by which the
accretion occurs are still not well understood, particularly for the most massive Herbig Be systems. Although it is generally agreed that a circumstellar 
accretion disk is present in these systems (for which there is direct observational evidence in a few cases), the structure and extent of the disks are
topics of active research and debate. 
 
MWC\,297 is a young ($\sim 0.1$ Myr), highly reddened (A$_V$ $\sim$ 8 mag), rapidly rotating ($v\sin i \sim 350$ km s$^{-1}$) early-type (B1.5 Ve) HAeBe star with a luminosity 
of $L > 3 \times$ 10$^3$~\Lsun\  \citep{Drew97, Benedettini01, Fairlamb15} at a distance of $417.9 \pm 5.3$~pc (based on Gaia; \citealt{Riello21}). The estimated distance places  
MWC\,297 in the Aquila Rift \citep{Ortiz-Leon17} and beyond the unrelated, foreground H {\small II}
region Sh 2-62 \citep{Drew97}. MWC 297 exhibits a
strong infrared excess due to hot dust in a circumstellar disk and surrounding envelope \citep{Hillenbrand92}.
It is one of the few early-type
B stars known to be surrounded by a circumstellar disk from which it is actively
accreting, as evidenced by strong and variable Balmer line emission and double-peaked CO emission
\citep{Finkenzeller84,Drew97,Banzatti22,Sandell22}. It was included in Herbig's original list of HAeBe
stars \citep{Herbig60} as well as in the comprehensive catalogue of HAeBe
members and candidate members compiled by \citet{The94}. 
Analysis of 2MASS and (saturated) {\it Spitzer}
IRAC images indicates that  MWC\,297 is the most massive member in a cluster of
$\sim$ 80 PMS stars \citep{Wang07, Gutermuth09}. Recent high-resolution near-infrared imaging observations have revealed the 
presence of two possible low-mass ($< 0.5 M_{\odot}$ companions within 250 AU of MWC 297 \citep{Ubeira20,Sallum21}.
MWC 297 has been observed across the electromagnetic spectrum and the literature on it is extensive. 
A summary of some of the results of previous observations can be found in \citet{Acke08}.

Because MWC 297 is so bright ($K \sim 3$ mag), with strong emission lines, and relatively nearby (for a massive star), it has frequently been the target of
near-infrared (NIR) and mid-infrared (MIR) interferometric observations. However, 
analyses of these, as well as other, observations of MWC 297 have led to contradictory results regarding the inclination $i$ of 
the accretion disk surrounding it. Based on the extended nature of the source in a 5 GHz radio map, as well as the high value of $v \sin i$ derived from the optical spectra,  \citet{Drew97} initially suggested that the inclination of the system was very high and seen nearly edge-on. The N-S elongation of the source seen in the radio maps of \citet{Sandell11} also suggests that the the disk is highly inclined (nearly edge-on with $i > 80$ \degr). However, the double-peaked optical [O I] emission lines detected by
\citet{Zickgraf03} are more indicative of an intermediate inclination for the system. On the other hand, the narrowness of the [O I] lines observed by 
\citet{Acke05} led to the conclusion that, if these lines arise from the bipolar outflow, as is typical of Herbig Be stars, then outflow axis should be in the plane of the sky and the system should be viewed nearly edge-on \citep{Acke08}.
Interpretation of the NIR, MIR, and mm interferometric results by numerous authors, however, indicates a fairly low inclination for the system ($i < 40$\degr and in some cases as low as $20$\degr; \citealt{Millan-Gabet01, Malbet07, Acke08, Alonso-Albi09, Weigelt11, Hone17, Kluska20}). 
Recently,  \citet{Sallum21} used the Large Binocular Telescope 
Interferometer to image MWC\,297 and derived an inclination
angle $\sim$ 50\degr\ - 65\degr\ under the assumption that the source consisted of a disk plus an
outflow at moderate inclination angles. 
Finally, \citet{Oudmaijer99} found no evidence 
for variation in the spectro-polarimetric signal across the H$\alpha$ line, a result which seems to indicate that the projection of the H $\alpha$ emission region in MWC 297 
onto the plane of the sky is circularly symmetric (that is, the system is seen nearly pole-on, with $i \approx 0$\degr). 
The discrepancy in the derived inclination angles is not merely a trivial detail, as the value can have profound consequences for our understanding of this object.
Given the $v \sin i$ value of 350 \kms found by \citet{Drew97}, an inclination $i < 40$\degr , would imply a stellar rotational velocity 
close to or even in excess of the critical, or break-up, velocity for such a massive star ($\sim 10$ M$_{\odot}$). 

In this paper we present mid-infrared images of MWC 297 obtained with the Faint Object infraRed
Camera for the SOFIA Telescope (FORCAST) on the Stratospheric Observatory
for Infrared Astronomy (SOFIA)\footnote{The NASA/DLR
Stratospheric Observatory for Infrared Astronomy (SOFIA) is jointly
operated by the Universities Space Research Association, Inc. (USRA), under
NASA contract NAS2-97001, and the Deutsches SOFIA Institut (DSI) under DLR
contract 50 OK 0901 to the University of Stuttgart.} as well as previously unpublished NRAO/Very Large Array (VLA) \footnote{The National Radio Astronomy Observatory is a facility of the National Science Foundation operated
under cooperative agreement by Associated Universities, Inc.} and Berkeley-Illinois-Maryland Association (BIMA) millimeter array observatory data. 
These observations show that the star drives an ionized jet surrounded by warm entrained dust. The morphology of
the outflow, as seen in these images, clearly rules out a face-on disk. We combine these data with {\it Herschel}\footnote{{\it Herschel}
is an ESA space observatory with science instruments provided by European-led Principal 
Investigator consortia and with important participation from NASA.} archival
data to construct the spectral energy distribution of the system and determine the total luminosity.
We also present a medium resolution (R $\sim$ 2000) 0.8 - 2.4 $\mu$m spectrum of MWC 297 obtained with SpeX at the IRTF
and use these data to better constrain the properties of the outflow (velocity and mass-loss rate) from the star.

We describe the observations in increasing order of wavelength 
in the next section. We then present results of attempts to reproduce the morphology seen in the VLA and FORCAST images
with a simple physical model. We find that the inclination of the system must be much larger than that derived from the interferometric 
measurements. We also present the results of a wind model constructed to determine the physical parameters of the outflow. A discussion 
of these results and our conclusions are given in the final section.  

\section{Observations and Data Reduction}

\subsection{IRTF/SpeX observations}
We observed  MWC\,297 at the NASA Infrared Telescope Facility (IRTF) on Mauna
Kea on 2007 Nov 02 (UT) with SpeX, the facility near-infrared medium resolution
cross-dispersed spectrograph \citep{Rayner03}. Ten individual exposure of MWC
297, each lasting 30 s, were obtained using the short-wavelength cross-dispersed
(SXD) mode of SpeX. This mode yields spectra spanning the wavelength range 0.8
-2.4 $\mu$m divided into 6 spectral orders. The observations were acquired in
``pair mode", in which the object was observed at two separate positions along
the 15\arcsec-long slit. The slit width was set to 0$\farcs$3, which yields a
nominal resolving power of 2000 for the SXD spectra. (At the distance of MWC\,
297 of 418 pc, the SpeX $0\farcs3$ slit spans $\sim 125$ au.) The slit was set
to the parallactic angle during the observations. The airmass was about 1.43 for
the observations. Observations of HD 171149, an A0$\, $V star, used as a
``telluric standard" to correct for absorption due to the Earth's atmosphere and
to flux calibrate the target spectra, were obtained immediately after the
observations of MWC 297. The airmass difference between the observations of the
object and the standard was 0.05. Conditions were reported to be rather poor
during the observations, with fog present. A set of internal flat fields and arc
frames were obtained immediately after the observations of MWC\, 297 for flat
fielding and wavelength calibration purposes.

\begin{figure*}
\includegraphics[scale=0.75, angle=90]{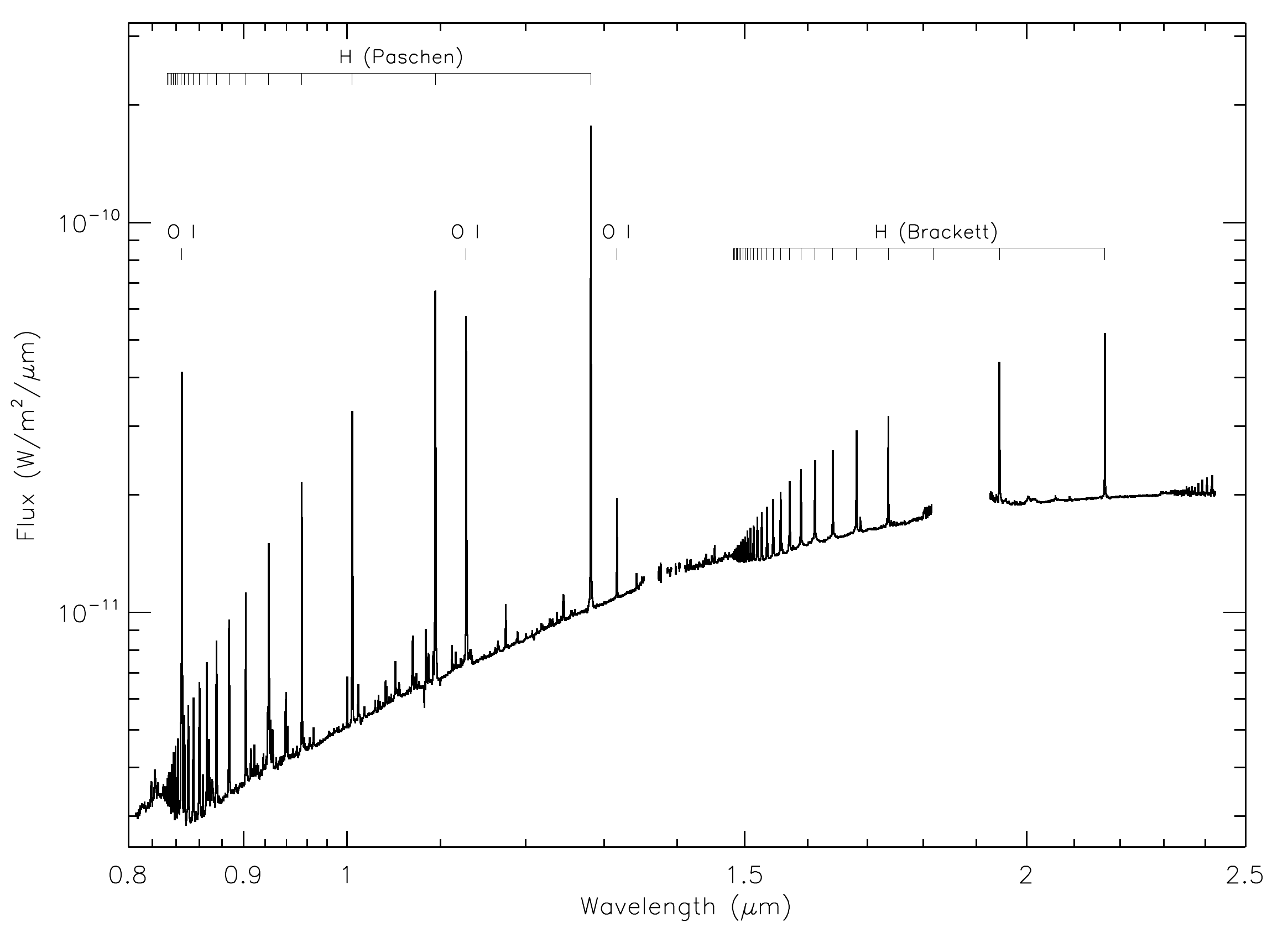}
\figcaption[]{
\label{fig-SpeX}
Spectrum of MWC 297 obtained with SpeX/IRTF. The Brackett and Paschen series of H have been
marked along with the prominent O I lines. Pfund lines, as well as a weak CO (2-0) emission feature, can 
be seen at the longest wavelengths $\lambda > 2.29 \mu$m). Continuum jumps associated with all three H series are also 
clearly seen. The weak, broad, double-peaked feature at $\sim 2 \mu$m is a residual telluric artifact. Aside
from the P Cygni shape of the He {\small I} 1.083 $\mu$m, no convincing evidence of (stellar) absorption lines 
can be seen.
}
\end{figure*}

The data were reduced using Spextool \citep{Cushing04}, the IDL-based package
developed for the reduction of SpeX data. The Spextool package performs
non-linearity corrections, flat fielding, image pair subtraction, aperture
definition, optimal extraction, and wavelength calibration. The sets of spectra
resulting from the individual exposures were median combined and then corrected
for telluric absorption and flux calibrated using the extracted A0\, V telluric
standard spectra and the technique and software described by \citet{Vacca03}.
The spectra from the individual orders were then spliced together by matching
the flux levels in the overlapping wavelength regions, and regions of poor
atmospheric transmission were removed. The final $0.8-2.4$ $\mu$m spectrum  is
shown in Figure~\ref{fig-SpeX}. The S/N varies across the spectrum but is on the
order of several hundred across the entire SXD wavelength range.

We computed synthetic NIR magnitudes in various filters from our final spectrum.
The estimated 2MASS magnitudes are $J = 6.31$, $H=4.61$, and $K_s=3.31$.
Comparison with the 2MASS point source catalogue indicates that our synthetic
magnitudes are systematically lower by $\sim0.22$ mag, most likely due to the
effects of losses through the narrow slit and the poor observing conditions.
However our synthetic $J-H$ and $H-K_s$ colors agree to better than 0.05 mags
with the 2MASS values, which indicates that our relative flux calibration is
quite accurate.\footnote{See also \citet{Rayner09}  where it is demonstrated that the adopted method
of flux calibration produces spectra with relative fluxes accurate to a few
percent.}

Fluxes for the strong emission lines seen in the spectrum were measured after
subtracting the continuum. We fitted the continuum between the Paschen, Brackett, and Pfund jumps 
seen in the spectrum with low order polynomials.
The continuum subtracted spectrum is shown in Fig. \ref{speccomp}. The line
fluxes are presented in Table \ref{SpeXfluxes}, and have been multiplied by 1.22 to account for
the necessary adjustment to the overall flux levels to match the 2MASS magnitudes (again, under
the assumption that the differences in the synthetic and 2MASS mags are due primarily to the 
poor observing conditions).

\subsection{SOFIA/FORCAST observations}

\begin{figure}
\hspace{1.25in}
\includegraphics[trim=1.5in 5in 0in 1in]{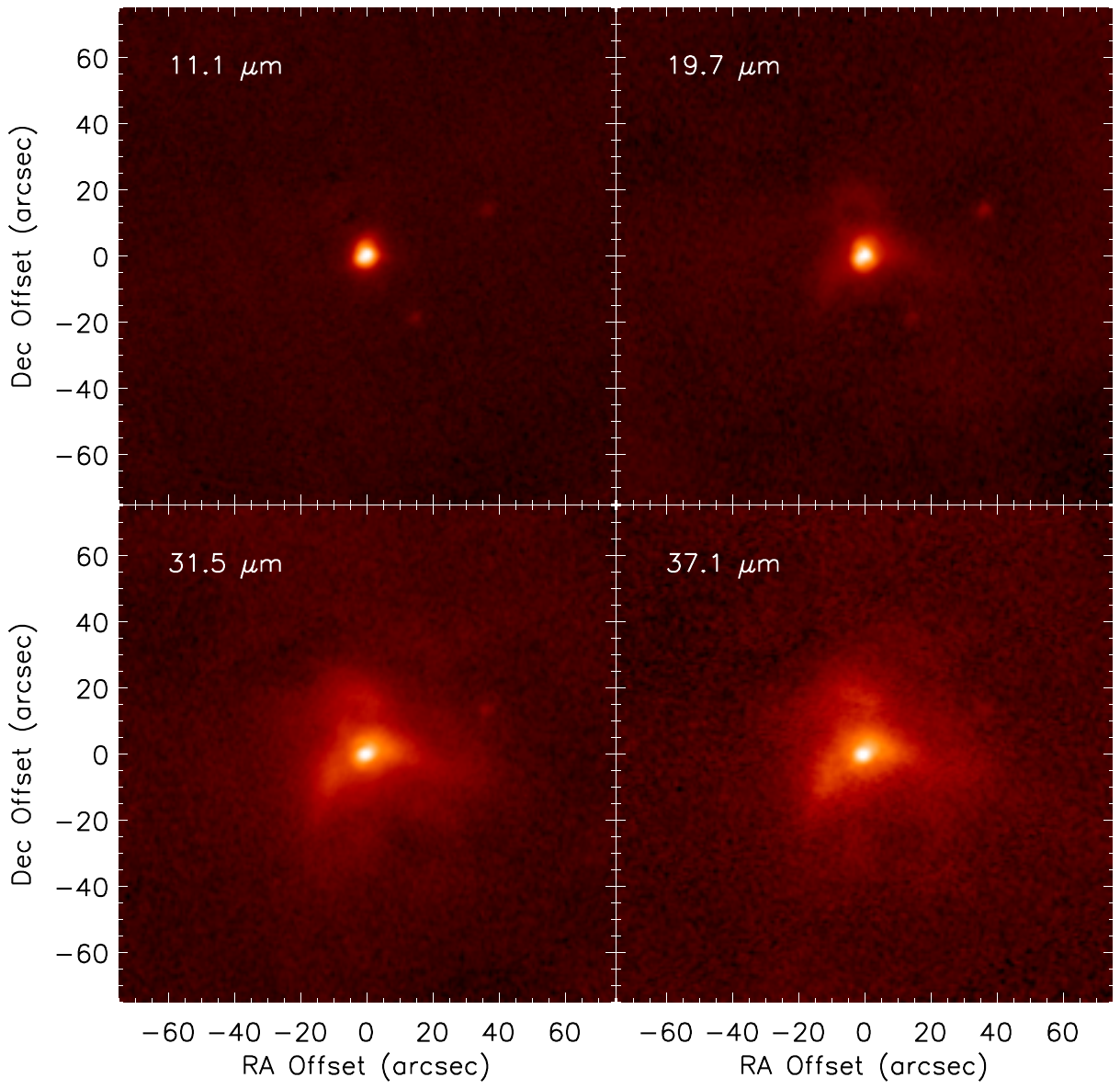}
\vspace{0in}
\figcaption[]{
\label{fig-forcast}
FORCAST images of MWC297 at 11, 19, 31, and 37 $\mu$m, displayed with asinh scaling \citep{Lupton99}. North is up and East is left. The two 
Class II sources MWC\,297\#1 and \#2 can be clearly seen at 11.1 and 19.7 $\mu$m. 
}
\end{figure}

\begin{figure}
\hspace{0.75in}
\includegraphics[scale=0.75,trim=0in 5in 0in 1in]{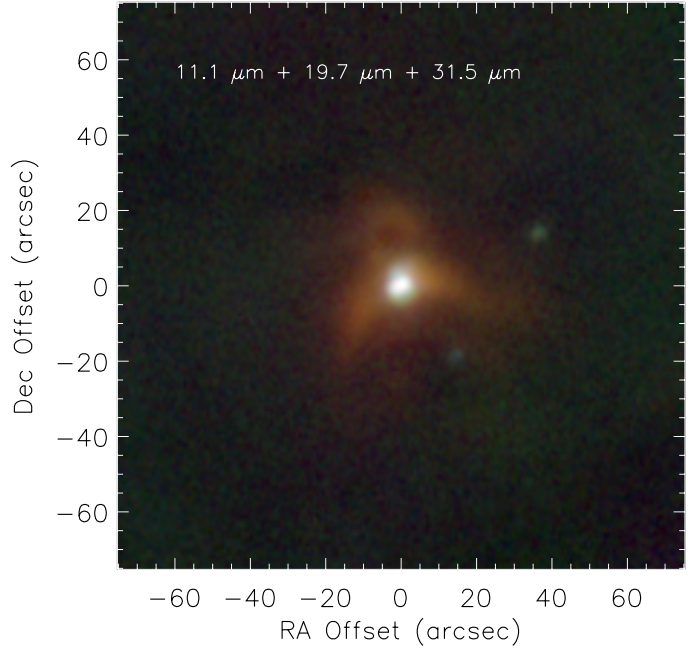}
\figcaption[]{
\label{fig-sofia-3color}
Three-color image generated by combining the FORCAST 11.1 (blue), 19.7 (green), and 31.5 (red) $\mu$m images.  }
\end{figure}

\begin{deluxetable*}{rclccccc}
\tabletypesize{\scriptsize}
\tablecolumns{8}
\tablenum{2}
\tablewidth{0pt} 
\tablecaption{SOFIA/FORCAST Observations of MWC\,297 \label{tbl-FORCASTobs}}
\tablehead{
\colhead{Filter} & \colhead{Flight} & \colhead{UT Date} & \colhead{UT Time\tablenotemark{a}} & \colhead{$t_{exp}$\tablenotemark{b}} & \colhead{Altitude}  & \colhead{Zenith Angle} & $C$\tablenotemark{c}  \\
\colhead{($\mu$m)} &    &      & & \colhead{(s)} & \colhead{(feet)} & \colhead{(deg)} & \colhead{(Me$^{-}$ s$^{-1}$ Jy$^{-1}$)} 
}
\startdata
6.6   & 170 & 2014 May 08  & 11:12:29 & 120 & 42983 & 43.6 & 0.093 $\pm$ 0.004 \\
11.1 & 170 & 2014 May 08 & 11:05:35  &  180 & 42973 & 44.0 & 0.363 $\pm$ 0.033 \\
19.7 & 170 & 2014 May 08 & 10:58:49  & 189 & 42993 & 42.8 & 0.530 $\pm$ 0.012 \\
31.5 & 170 & 2014 May 08 & 11:05:35  &  222 &  42973 & 44.0 & 0.165 $\pm$ 0.013\\
37.1 & 170 & 2014 May 08 & 10:58:49 &  429 & 42993 & 42.8 & 0.044 $\pm$ 0.004 \\
 & & & & & & & \\
11.1 & 178 & 2014 Jun 11 & 08:25:21 & 180 & 43001 & 43.6 & 0.354 $\pm$ 0.010 \\
19.7 & 178 & 2014 Jun 11 & 08:12:15 & 208 & 42995 & 44.6 & 0.535 $\pm$ 0.016 \\
31.5 & 178 & 2014 Jun 11 & 08:25:21 & 197 &  43001 & 43.6 & 0.106 $\pm$ 0.004\\
37.1 & 178 & 2014 Jun 11 & 08:12:15 & 199 & 42995 & 44.6 & 0.028 $\pm$ 0.003 \\
\enddata
\tablenotetext{a}{UT Time, Altitude, and Zenith Angle refer to the start of the observations}
\tablenotetext{b}{On-source exposure time}
\tablenotetext{c}{Calibration factor, and associated uncertainty, used to convert count rates in reduced images from Mega-electrons s$^{-1}$ to Jy; see \citep{Herter13, Clarke21}}
\end{deluxetable*}

\begin{deluxetable*}{lllrrccc}
\tabletypesize{\scriptsize}
\tablecolumns{8}
\tablenum{3}
\tablewidth{0pt} 
\tablecaption{Positions and flux densities of FORCAST mid-infrared  sources in the MWC\,297 field\label{tbl-FORCAST}}
\tablehead{
\colhead{Source} & \colhead{$\alpha$(2000.0)} & \colhead{$\delta$(2000.0)}  & \colhead{S(6.6 $\mu$m)}& \colhead{S(11.1 $\mu$m)} & \colhead{S(19.7 $\mu$m)} & \colhead{S(31.5 $\mu$m)} & \colhead{S(37.0 $\mu$m)}   \\ 
   & \colhead{[$^h$  $^m$ $^s$]}& \colhead{[$^\circ$ \arcmin\ \arcsec ]}& \colhead{[Jy]} & \colhead{[Jy]}  & \colhead{[Jy]} & \colhead{[Jy]}  & \colhead{[Jy]} 
}
\startdata
MWC\,297 core & 18 27 39.53  & $-$03 49 52.1 & 109.9  $\pm$ 4.8 & 118.8 $\pm$ 3.5  & 109.5 $\pm$ 3.4 & 113.9 $\pm$ 15.4  & 97.7 $\pm$ 18.6  \\
\#1\tablenotemark{a}& 18 27 37.07 & $-$03 49 38.5 & 0.88  $\pm$ 0.02 &  1.04 $\pm$ 0.01  & 1.47  $\pm$ 0.12 & \nodata     & \nodata  \\  
\#2\tablenotemark{b}& 18 27 38.55  & $-$03 50 11.0 & 1.04  $\pm$ 0.01 & 0.93 $\pm$ 0.01 & 0.67 $\pm$ 0.07 & \nodata  & \nodata \\	
\#3& 18 27 38.69  & $-$03 49 54.1 & 0.43  $\pm$ 0.02   &  $<$ 0.06  & \nodata   & \nodata & \nodata  \\
\#4 & 18 28 38.17 & $-$03 49 44.0 &  0.80  $\pm$ 0.02  &   0.34 $\pm$ 0.03 & \nodata & \nodata  & \nodata  \\
\#5 & 18 27 38.83  & $-$03 49 34.8 & 0.61  $\pm$ 0.02  &  0.44 $\pm$ 0.03 & \nodata  & \nodata & \nodata  \\
\#6 & 18 27 38.95 & $-$03 50 01.4 & 0.22 $\pm$ 0.03 &   $ <$ 0.06  & \nodata & \nodata  & \nodata  \\
MWC\,297 extd\tablenotemark{c} & 18 27 39.53  & $-$03 49 52.1 & 111.4  $\pm$ 4.9& 121.6 $\pm$ 3.6  & 171.6 $\pm$ 5.3 & 513.1 $\pm$ 18.5  & 650.4 $\pm$ 64.7  \\
 \enddata
 \tablenotetext{a}{2MASS J18273709-0349385, [GMM2009] MWC297 6 }
 \tablenotetext{b}{2MASS J18273854-0350108, [GMM2009] MWC297 7}
 \tablenotetext{c}{extended region surrounding MWC 297}
\end{deluxetable*}

MWC\,297 was observed  with FORCAST on SOFIA as part of a cycle 2
project (02\_0016) targeting disks around early B-stars. FORCAST is a
dual-channel mid-infrared camera covering 5 - 40 $\mu$m with a suite of broad
and narrow-band filters \citep{Herter12}. Each channel consists of an array of
256 $\times$ 256 pixels. Both channels can be operated simultaneously via a
dichroic mirror internal to FORCAST, which was the configuration used for these
observations. MWC\,297 was observed using the `Nod-Matched-Chop' mode with
dithering. On 08 May 2014, all observations were (erroneously) executed with a
30\arcsec\ east-west chop throw and nod, using the filter combinations 19.7/37.1
$\mu$m, 11.1/31.5 $\mu$m, and 6.6/37.1 $\mu$m, 
resulting in twice the
integration time for 37.1 $\mu$m. Unfortunately, because of the the large
angular extent of the emission in MWC297, the relatively small chop/nod throw
renders these images useful only for investigating point sources in the vicinity
of MWC297. The observations were repeated on 11 June 2014 with a 270\arcsec\
chop throw at a position angle of 320\degr, in the filter combinations 19.7/37.1
and 11.1/31.5 $\mu$m. These images are shown in Figure~\ref{fig-forcast} and were used
to measure the flux from MWC297 and for the subsequent analysis.
On both flights MWC\,297 was observed at an aircraft altitude of $\sim 43000$ feet with
excellent sky transmission, although the sky background was somewhat variable on
the second flight. Details of the observations are given in Table \ref{tbl-FORCASTobs}.

We retrieved the Level 3 data from the SOFIA Science archive. These pipeline
reduced and calibrated images require no further processing; details about
FORCAST data reduction and calibration can be found in \citet{Herter13}. The
pixel size after pipeline processing is 0.768\arcsec. We found that the relative
positions of the source were different in the various filters. We manually
shifted the images such that the centroid of the peak of the emission in all the
filter images was aligned with that seen in the 11.1 $\mu$m filter.

Figure~\ref{fig-sofia-3color} shows a three color image of MWC\,297 generated from the
11.1, 19.7, and 31.5 $\mu$m filter images obtained during the second SOFIA
flight. At wavelengths shorter than  19.7 $\mu$m, the images reveal only unresolved
point sources, and the data from the first flight can also be used for point source
photometry. At 6.6
$\mu$m we see six sources, three of which are new detections. All the new
sources are within 20\arcsec\  of MWC\,297, which is severely saturated in
Spitzer and WISE images. This  makes it impossible to see even a moderately faint
source in the vicinity of the star. At 11.1 $\mu$m and longer wavelengths only
MWC\,297 and the two brightest young stellar objects are visible. The latter two
both have 2MASS counterparts and were first detected in the mid-infrared by
\citet{Habart03} at 10.5 $\mu$m\footnote{Unfortunately \citet{Habart03} reversed the RA offsets
measured from MWC\,297. The RA coordinates for MWC\,297\#1 and \#2 are therefore
incorrect in the SIMBAD data base.}.  MWC\,297\#1 is also a sub-millimeter source,
first detected by \citet{Henning98} and is also visible in the SCUBA 850
and 450 $\mu$m images presented by \citet{Sandell11}. In \citet{Gutermuth09},
who observed the MWC\,297 region with {\it Spitzer} IRAC and MIPS 
they are listed as MWC\,297 \#6 and \#7. They characterize both as Class II
objects.

Flux densities from PSF fitting or aperture photometry (using a radius of 12 pixels) of
the central stellar core are presented in
Table~\ref{tbl-FORCAST}, along with measurements for the other point sources detected in the FORCAST images. 
As can be seen in Figures~\ref{fig-forcast} and \ref{fig-sofia-3color}, at wavelengths longer than
11.1 $\mu$m, the emission from the envelope surrounding the central star in MWC297 is
quite prominent. Since we expect that this emission arises from dust reprocessing the ultraviolet
and optical light from the central source, we used a large aperture with a radius of $54$\arcsec 
(70 pixels) to measure the total flux from the extended MWC297 region (stellar core plus envelope) in these images.

\subsection{{\it Herschel} PACS and SPIRE archival data}

The MWC\,297 field was observed twice in the guaranteed time Key
Program the {\it Herschel} Gould Belt Survey (HGBS)
\citep{Andre10,Bontemps10,Konyves10}. On Operational Day (OD) 163 (2009-10-24)
MWC\,297 was observed in parallel mode with PACS (70/160 $\mu$m) and SPIRE.
The field was observed with fast scanning (60\arcsec{}/sec) with an
orthogonal scan, resulting in an approximately symmetric, but significantly
broadened PSF due to the fast scan rate. The PACS 70 $\mu$m data were
retrieved from the HGBS website. These data are  part of the observations
discussed in \citet{Bontemps10} as well as in \citet{Konyves10, Konyves15}. We also retrieved the level 2.5 SPIRE data
from the {\it Herschel} data archive. MWC\,297 was
observed  on OD 522 (2010-10-18) with PACS (100/160 $\mu$m) with medium
scan speed (20\arcsec{} s$^{-1}$) with an orthogonal cross scan.  The latter
data set has much better image quality. Level 2.5 JScanam images processed
with HIPE 12.1.0 were retrieved for both 100 $\mu$m  and 160 $\mu$m from
the  {\it Herschel} data archive. 

\begin{figure*}
\hspace{1.25in}
\includegraphics[scale=0.48,angle=90]{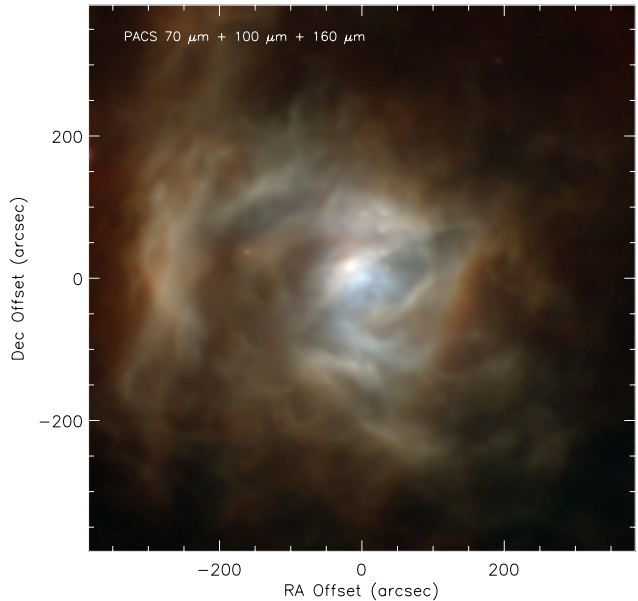}
\figcaption[]{
\label{fig-PACS_3color}
Three-color image of of MWC 297, generated from the PACS 70 (blue), 100 (green), and 160 (red) $\mu$m data on an asinh scale, of the region
used to estimate total MIR fluxes. The arc detected in the SOFIA/FORCAST images, as well as the NE core, offset by about 15 \arcsec from the center of MWC297,
can be clearly seen. The arc is blue while the NE core is considerably redder. A faint blue halo can also be seen centered on the NE core. The dust ridge
separating the central star in MWC297 from the NE core can also be seen.
}
\end{figure*}

\begin{figure}
\vspace{-0.15in}
\hspace{2.0in}
\includegraphics[ scale=0.5,trim=1in 1in 1in 4.0in]{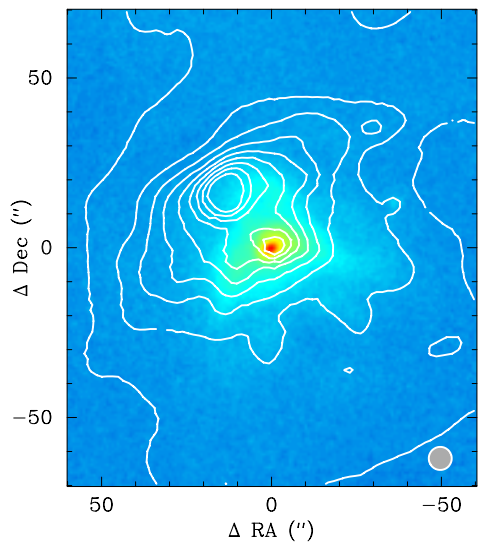}
\figcaption[]{
\label{fig-PACS}
PACS 100 $\mu$m image (contours) overlaid on a logarithmically stretched FORCAST 31.5 $\mu$m image in color.
}
\end{figure}

\begin{figure}
\hspace{2.0in}
\includegraphics[scale=0.65]{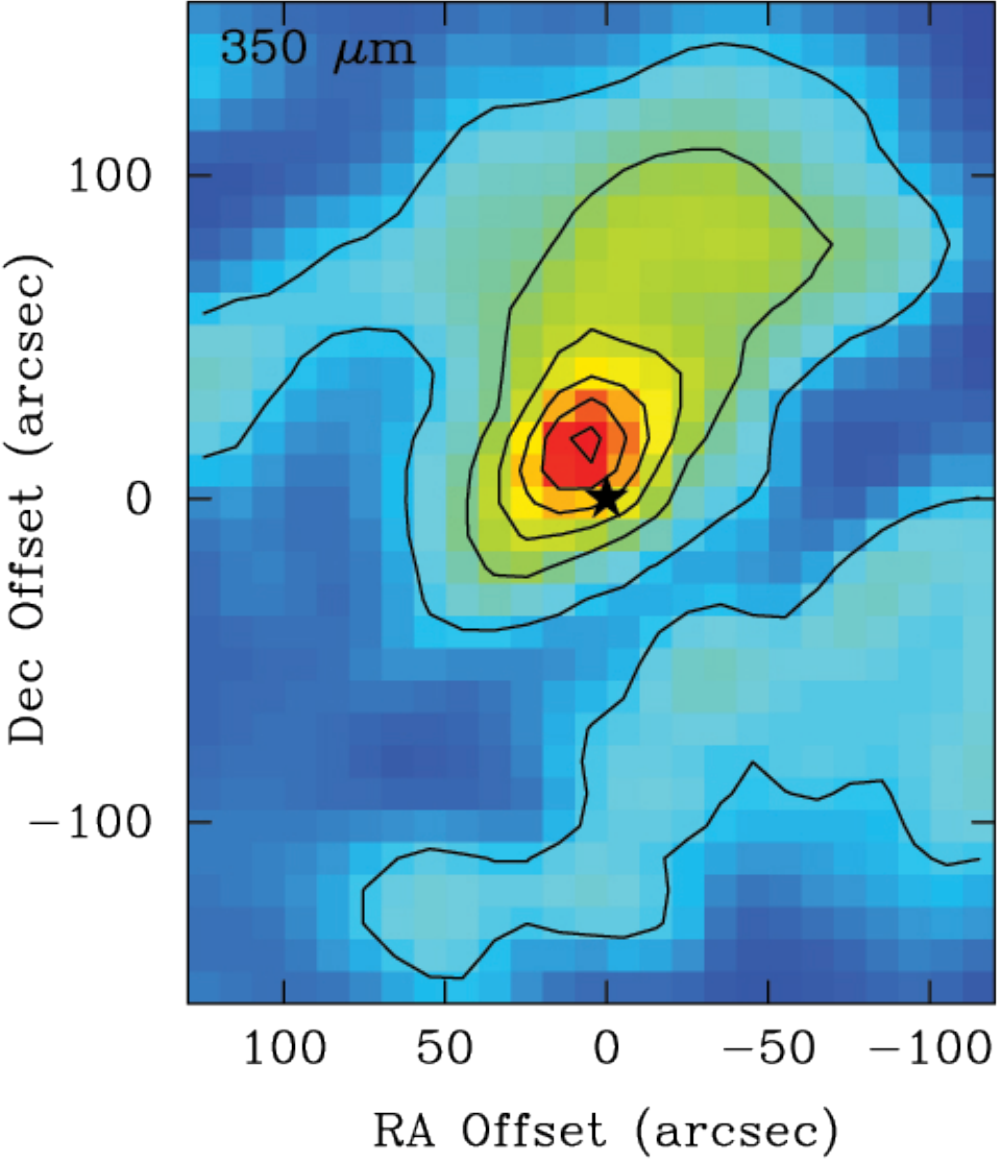}
\figcaption[]{
\label{fig-350um}
The 350 $\mu$m SPIRE image shows that MWC\,297 is located on the southwestern rim of
a small cloud. The emission at 350 $\mu$m is dominated by a prestellar core about 15\arcsec\ 
northeast of the star. The gap to the southwest of the star is also seen in the $^{13}$CO(1 -- 0)
image by \citet{Ridge03}
}
\end{figure}

Even though there is extensive emission
surrounding MWC\,297 from the molecular cloud associated with it, the
central star is clearly visible in the PACS images (Fig. \ref{fig-PACS_3color}). The emission from MWC\,297 is
particularly obvious in the PACS 100 $\mu$m image, which has the best spatial
resolution ($\sim$ 6.7\arcsec) of the three PACS images we analyzed. This is shown in 
Figure~\ref{fig-PACS}, where we overlay
the PACS 100 $\mu$m image on the FORCAST 31.5 $\mu$m image.  However, the central source cannot be seen in the SPIRE images,
which are completely dominated by the molecular cloud emission (see Fig. \ref{fig-350um}). Note that the brightest source
in the PACS 100 and 160 $\mu$m images and the SPIRE images is not the central star in MWC297, but rather
a source about $20$\arcsec ~ to the northeast, which we refer to as the NE core. This object sits on the edge of the 
diagonal dust ridge and seems to be encircled by a faint blue halo. It can be clearly seen in the 450 and 850 $\mu$m 
maps shown by \citet{Alonso-Albi09} and \citet{Sandell11}.
We also detect  MWC\,297\#1
at 100 $\mu$m and  at 160 $\mu$m, but here the source is too blended with
emission from the surrounding cloud to allow us to determine a reliable flux density.  MWC\,297\#1
is not detected at 70 $\mu$m. 

A three-color image generated by combining the
PACS 70, 100, and 160 $\mu$m data is shown in Figure~\ref{fig-PACS_3color}. In addition to measuring the flux from the 
central source in the three PACS bands, we also used a large aperture with a radius of $384$\arcsec
(120 pixels) in order to capture all of the flux from the envelope surrounding the central star, which 
is expected to be reprocessed light from the central star. A similar sized aperture was used to estimate
the fluxes in the SPIRE images. The fluxes are given in Table~\ref{tbl-PACS}.

\begin{deluxetable*}{lllcrcl}
\tabletypesize{\scriptsize}
\tablecolumns{7}
\tablenum{4}
\tablewidth{0pt}
\tablecaption{Positions and flux densities of Herschel PACS  sources in
  the vicinity of MWC\,297.\label{tbl-PACS}}
\tablehead{
Name & \colhead{$\alpha$(2000.0)} & \colhead{$\delta$(2000.0)}  & \colhead{S(70 $\mu$m)}& \colhead{S(100 $\mu$m)} & \colhead{S(160 $\mu$m)}  & comment \\
\colhead{}  & \colhead{[h m s]}& \colhead{[$^\circ$ \arcmin\ \arcsec ]}& \colhead{[Jy]} & \colhead{[Jy]}  & \colhead{[Jy]}   & \\
}
\startdata
MWC\,297  core   & 18 27 39.51 &  $-$03 49 51.0 & 17.8 $\pm$ 5.0 & 16.3 $\pm$ 3.3  & 7.0 $\pm$ 3.0 & \\
MWC\,297\#1      & 18 27 37.31 & $-$03 49 38.7 & \nodata              & 1.8 $\pm$ 0.3    & \nodata &  \\
NE Core              & 18 27 40.56 & $-$03 49 34.7 & 51.7 $\pm$ 15.5  & 77.4 $\pm$ 8.0  & 45.6 $\pm$ 5.0  & [KAM2015]68\\
MWC\,297  extd   & 18 27 39.51 &  $-$03 49 51.0 & 4838 $\pm$ 160 & 6557 $\pm$ 230  & 6142 $\pm$ 441 & \\
\enddata
\end{deluxetable*}

\subsection{BIMA observations}

\begin{figure}[h]
\hspace{1.7in}
\includegraphics[width=0.45\textwidth]{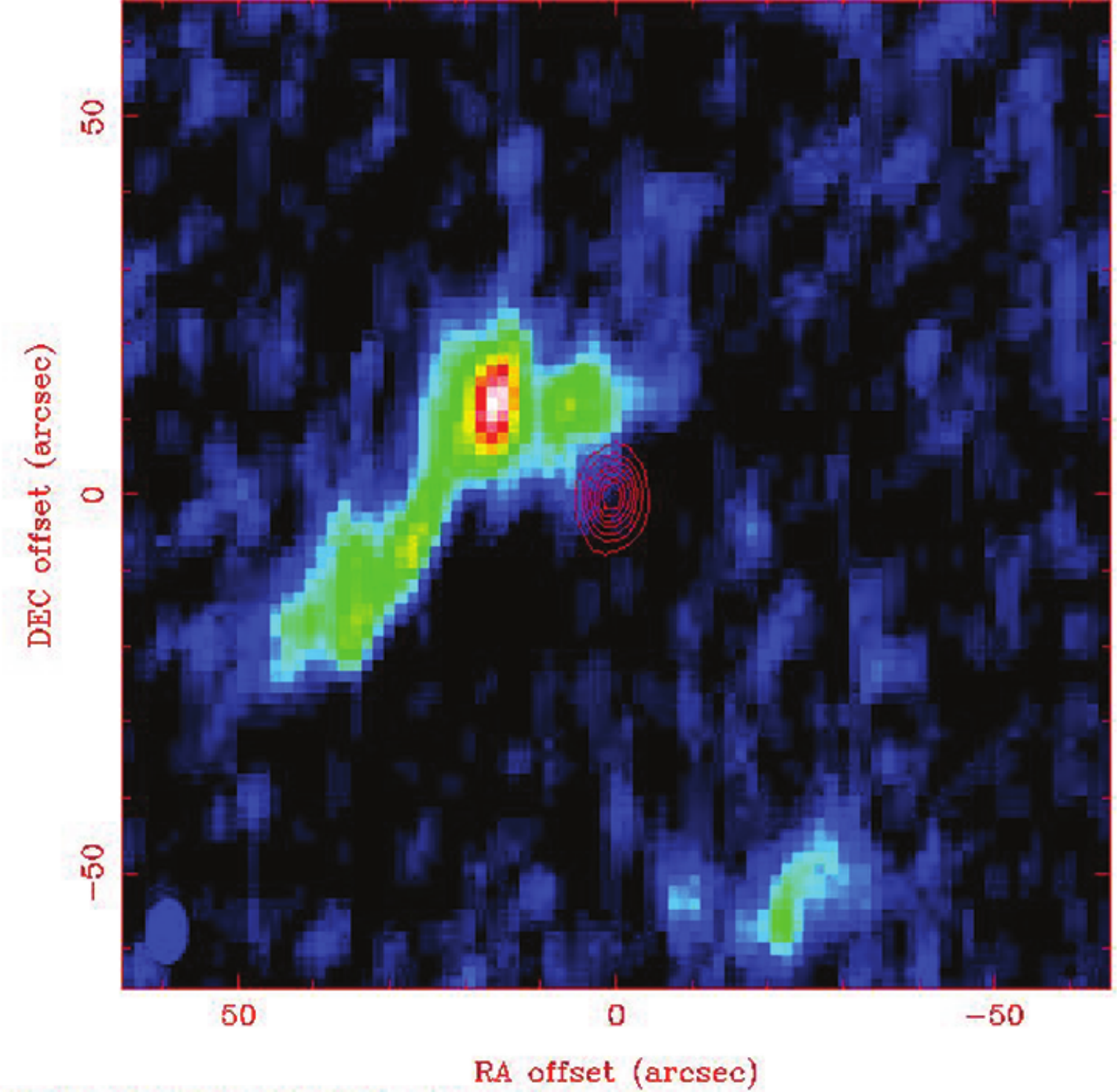}
\caption{Integrated $^{13}$CO(1--0) emission plotted in color with 110 GHz continuum emission overlaid with red contours
\label{fig-bima}}

\end{figure}

Observations of MWC\,297 were conducted on several occasions between
2002 October  and 2003 February  at several frequencies and antenna
configurations; see Table~\ref{tbl-bima}. The mm-emission is unresolved in all
observations with the possible exception of the B-array data obtained on 2003
February 18. Therefore the synthesized beam size is not listed in Table 2
except for these 2003 February data. All data sets
were reduced and imaged in a standard way using MIRIAD \citep{Sault95}. The
quasar 1751$+$096 was used for phase calibration, 3C454.3 was used for amplitude
calibration, and Uranus or MWC\,349 was used for flux calibration. Since the primary
objective of these observations was to characterize the continuum emission
from MWC\,297, the correlator was configured for four 100 MHz bands, except
for the 110 GHz observations where  $^{13}$CO(1--0) was centered in a 25
MHz wide window, while the other three correlator windows were set to 100
MHz. Figure \ref{fig-bima} shows the B-array image with the integrated $^{13}$CO(1--0)
emission in color overlaid with continuum emission in contours.

\begin{deluxetable}{cllc}
\tabletypesize{\scriptsize}
\tablecolumns{4}
\tablenum{5}
\tablewidth{0pt} 
\tablecaption{BIMA flux densities of MWC\,297\label{tbl-bima}} 

\tablehead{
\colhead{Frequency}   &  \colhead{Flux density} & \colhead{Observing date(s)} & \colhead{Array configuration}\\
\colhead{[GHz]} &  \colhead{[mJy]} & & \\
}
\startdata
\phantom{0}75  & \phantom{0}85 $\pm$ 0.15  &  2002 Nov 29 & C \\
107  & 149.6   $\pm$ 0.08 & 2002 Oct  28 & C\\
108.5\tablenotemark{a}  & 146.25 $\pm$ 0.11 & 2003 Feb 17 & B\\
230   & 340 $\pm$  0.15 &  2002 Nov 26, 29 & C \\
\enddata
\tablenotetext{a}{average of upper and lower sideband, synthesized beam 3\ptsec62 $\times$ 2\ptsec08, PA = 10\degr{}}
\end{deluxetable}

\subsection{VLA observations and VLA archival data}

\begin{figure}
\vspace{-1.5in}
\hspace{1.0in}
\includegraphics[width=5in, angle=-90, trim=1.0in 0.5in 1in 5in]{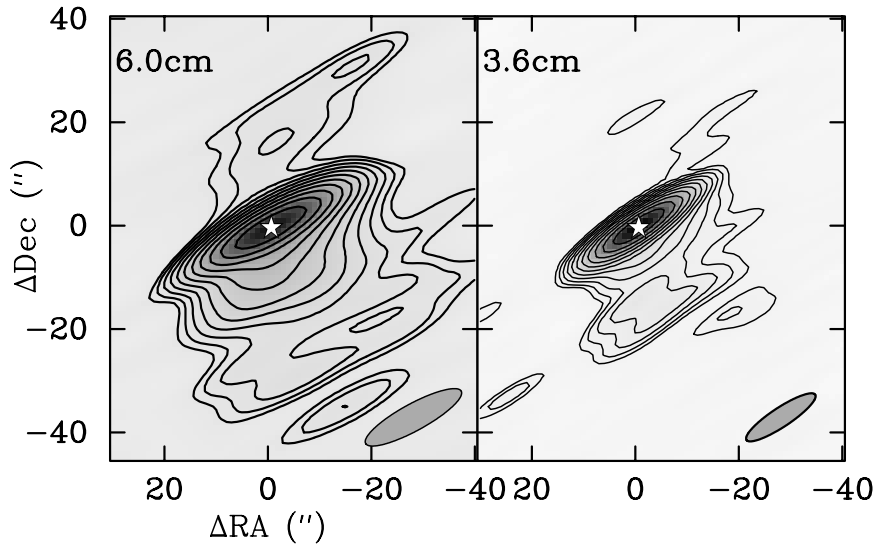}
\figcaption[]{
\label{fig-vla-cnd}
Re-reduced VLA CnD observations at 6.0  and 3.6 cm. The beam,
which is very elliptical, is drawn in the right bottom corner of each
image.
}
\end{figure}

MWC\,297 was observed with the VLA on 2003 December 5 in the C (4.89
GHz) and U-band (14.94 GHz) for 4.5 hrs
in the B-configuration. The observing conditions were excellent and we had good UV
coverage, giving a synthesized beam of 1.55\arcsec\ $\times$ 1.15\arcsec\ with a 
PA $= -22.1$\degr\  in C-band and 0.49\arcsec\ $\times$ 0.40\arcsec\, with a PA $= -15.1$\degr\ in U-band. 3C 286 was used for absolute calibration and
1832$-$105 for phase calibration. The data were reduced with AIPS in a
standard way. MWC\,297 was bright enough for self
calibration, but the phase stability was so good that we could see no
improvement in signal to noise with this option. We resolved the radio emission at both
frequencies; see Table~\ref{tbl-vla}.

We searched the VLA archive and found a set of observations from 
1978 December 9 - 10 (L and C band), which we retrieved, reduced,
calibrated and created cleaned images at both frequencies.  At the time VLA
 had only 9 antennas working, mostly on the east and the west arms. The
bandwidth was 12.5 MHz and data were recorded for both R and L
polarization. The maximum antenna spacing was 14.4 km, which places the
antenna configuration somewhere between A and B.  At L-band (1.48 GHz) the
emission is extended with an integrated flux density of  8.8 $\pm$ 2.0 mJy.
 The C-band observations had better phase stability than the L-band
observations and resulted in an rms noise of 0.4  mJy beam$^{-1}$ with a
synthesized beam width of 3\farcs62 $\times$ 2\farcs21  and a PA = $-$19\degr. We
find MWC\,297 unresolved with a  total flux density of 5.1 $\pm$ 0.4 mJy,
which agrees quite well with our results from 2003, see Table~\ref{tbl-vla}.

Our observed flux densities agree quite well with those of 
\citet{Skinner93}, except for their CnD (low spatial resolution)
observations from 1991 Feb 7. We retrieved the latter data set from the VLA
archive and re-reduced the data. The higher flux densities
in the CnD data results from the inclusion of the bipolar ionized outflow lobes
from MWC\,297 (Fig. \ref{fig-vla-cnd}, which are filtered out in in the A and
B-array data. Performing a careful two-component Gaussian fit (one for the
compact core and one for the extended emission), we find very consistent
results for the core emission (see Table~\ref{tbl-vla} and \citealt{Skinner93}). The flux
densities for the outflow lobes were estimated by integrating over the role
are of emission and subtracting out the flux densities for the compact
core. The CnD observation also included L-band (1.49 GHz) observations of
the MWC\,297 field. Here the emission is dominated by the outflow lobes and
an estimate of the flux of the compact core could not be obtained. The total flux density
flux at 1.49 GHz (core + outflow lobes) is 15.3 $\pm$ 2.1 mJy.

\begin{deluxetable*}{lcccccc}
\tabletypesize{\scriptsize}
\tablecolumns{7}
\tablenum{6}
\tablewidth{0pt} 
\tablecaption{VLA data on MWC\,297 \label{tbl-vla}} 

\tablehead{
\colhead{Frequency}  & \colhead{Core size} & \colhead{PA}& \colhead{rms} & \multicolumn{3}{c}{Flux Density}\\
\colhead{}                    & \colhead{}          & \colhead{}                        & \colhead{}          & \colhead {Peak} & \colhead{Core Total}            & \colhead{Outflow Lobes} \\
\colhead{[GHz]} & \colhead{\arcsec{} $\times$ \arcsec{}} & \colhead{[deg]}  & \colhead{[mJy~beam$^{-1}$]} & \colhead{[mJy]} & \colhead{[mJy]} & \colhead{[mJy]}
}
\startdata
\phantom{0}4.86\tablenotemark{a} &     \nodata  & \nodata      & 0.074   & \phantom{0}7.04   $\pm$ 0.07  & 7.48 $\pm$ 0.01  & 8.6 \\
\phantom{0}4.89 &  0.33 $\times$ 0.15 & \phantom{0}$-$90 $\pm$ 12 &  0.034  & \phantom{0}6.18 $\pm$ 0.03  & 6.56 $\pm$  0.06& \nodata \\
\phantom{0}4.9\phantom{0}\tablenotemark{b}  & 1.3 $\times$ 0.8 & \phantom{0}$+$160 $\pm$ 18 & 0.17\phantom{0} & 5.1 $\pm$ 0.4 &5.1 $\pm$ 0.4& \nodata \\ 
\phantom{0}8.44\tablenotemark{a}  & \nodata & \nodata & 0.052 &  10.67  $\pm$ 0.05   &  10.82 $\pm$ 0.10      &  5.2  \\
\phantom{0}8.44\tablenotemark{c}  & \nodata & \nodata & 0.046 & 6.35     $\pm$ 0.06   &  8.88 $\pm$ 0.10      &  \nodata \\
\phantom{0}8.44\tablenotemark{c}  & \nodata & \nodata & 0.046 & 6.19     $\pm$ 0.07   &  8.80 $\pm$ 0.10      &  \nodata \\
14.93                   &  0.14 $\times$ 0.11 & \phantom{0}$+$26 $\pm$ 16            &  0.067 & 17.79 $\pm$  0.07 & 19.62 $\pm$ 0.12 & \nodata \\
\enddata
\tablenotetext{a} {re-reduced CnD data from \citet{Skinner93}}
\tablenotetext{b} {Archive data from Dec 10, 1978, see text}
\tablenotetext{c} {A/D ($\sim$ A) data from \citet{Skinner93}}
\label{VLAfluxes}
\end{deluxetable*}

\section{Results}

\subsection{Morphology}

The {\it Herschel} PACS and SPIRE images show that MWC\,297 lies on the southeastern 
rim of a ridge with a more extended cloud core north of the star. There
is a gap or low density area parallel to the ridge seemingly splitting the
cloud into a northern and a southern or southwestern core, the latter being less
dense (Figure~\ref{fig-350um}). The same structure is also seen in the 
SCUBA-2 images \citet{Rumble15}, although the cloud to the southwest is rather
faint. Maps of  $^{13}$CO  J = 1-- 0  and C$^{18}$O J = 1-- 0 \citep{Ridge03}
show the same  southeastern ridge as seen in the PACS images, including the
cloud  southwest of  MWC\,297. The $^{13}$CO  map shows that there is
essentially a cavity southwest of the ridge splitting the clouds in two. In the
sub-millimeter images one can see a protostellar core northeast of the star,
which starts to dominate the emission shortward of 850 $\mu$m. At 350 $\mu$m it
is very bright and MWC\,297 is no longer visible.  This core is also associated
with strong  $^{13}$CO emission seen in the  BIMA image (Figure~\ref{fig-bima}). 

The VLA data obtained 
by \citet{Skinner93}, and re-reduced by \citet{Sandell11}, in  the CnD  configuration show that MWC\,297 powers a large bipolar
ionized jet \citep[see also][]{Rumble15}, or outflow, oriented
approximately north-south. In the south the outflow has a wide opening angle,
while in the north it appears much weaker and narrower. The thermal emission is
dominated by a compact core, centered on the star, with a
spectral index, $\alpha \sim$ 1.03 and a size decreasing with increasing
frequency, as typically seen in collimated thermal jets
\citep{Reynolds86,Sandell09}. Sub-arcsecond imaging at 5 GHz \citep{Drew97}
resolves the core into a double peaked east-west structure (presumably the
ionized disk) with a more pronounced north-south extension in the outflow direction. 

In the MIR, at wavelengths $\lambda < 19\mu$m, MWC\,297 appears as an unresolved, point-like
source in the FORCAST images. However, at longer MIR wavelengths, the source is
clearly extended. An wide arc, opening to the southwest, whose tips are separated by $\sim 32$\arcsec
($\sim 13400$ au), is clearly seen in the 19, 31, and 35 $\mu$m images. The arms of the arc extend about $\sim
14 - 18$\arcsec (5850 - 7520 au) from the central source. 
A small loop, or annulus, with a diameter of $\sim
10$\arcsec (4180 au), can be seen to the northeast in the 19 and 31 $\mu$m images. The annulus and arc are
very nearly symmetric with respect to the central source, with the arc opening in nearly the opposite direction to 
the position of the northern annulus. 

\begin{figure}
\hspace{1.6in}
\vspace{0.175in}
\includegraphics[scale=0.6, trim= 0.75in 0.75in 0.75in 4in]{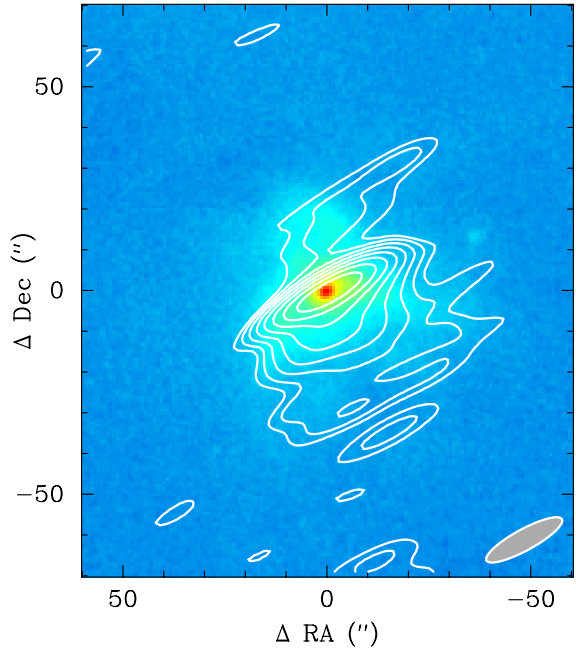}
\figcaption[]{
\label{fig-vla-SOFIA}
VLA 6 cm  image (contours) overlaid on a logarithmically stretched FORCAST 31.5 $\mu$m image in color.
The VLA beam is plotted at bottom right.}
\end{figure}

The structure seen in the long wavelength FORCAST images mirrors that seen in the VLA data.
A comparison between the FORCAST 31 $\mu$m image and the VLA 6 cm map is shown in Figure \ref{fig-vla-SOFIA}.
These data suggest that the ionized outflow from MWC 297 is surrounded by a warm dust layer, which is perhaps being
compressed and heated in the shear layer between the outflow and the surrounding molecular cloud. The compact nature of the emission
to the north indicates the the outflow is constrained by the dense environment in this direction. 
Therefore, we interpret the arc and the northern loop seen in the FORCAST images as arising from 
dust on a bi-conical surface, surrounding the outflow, that is tipped with respect to the line of sight; the southwestern
arc traces the edge-brightened surface that is tipped toward us and the northeastern annulus traces the circular rim 
of the northern structure that is tipped away from us. Diagrams of similar outflow structures can be seen in the papers by 
\citet{Canto81, Raga93}.
\footnote{The referee has asked us to consider whether the arc could be 
a bow shock. Given the symmetry of the structures seen in the MIR images and radio maps, the fact that the MIR structures align well with the 
ionized outflows seen in the radio maps, and the lack of any nearby source that would produce a wind strong enough to create a
bow shock, we do not consider this to be likely. A simpler and more likely explanation is that the structures seen in the MIR images
are produced by emission from dust in the outflow cavity walls.}

\begin{figure*}
\begin{minipage}{0.5\textwidth}
\hspace{-0.1in}
\vspace{1.25in}
{\includegraphics[scale=0.4,angle=90]{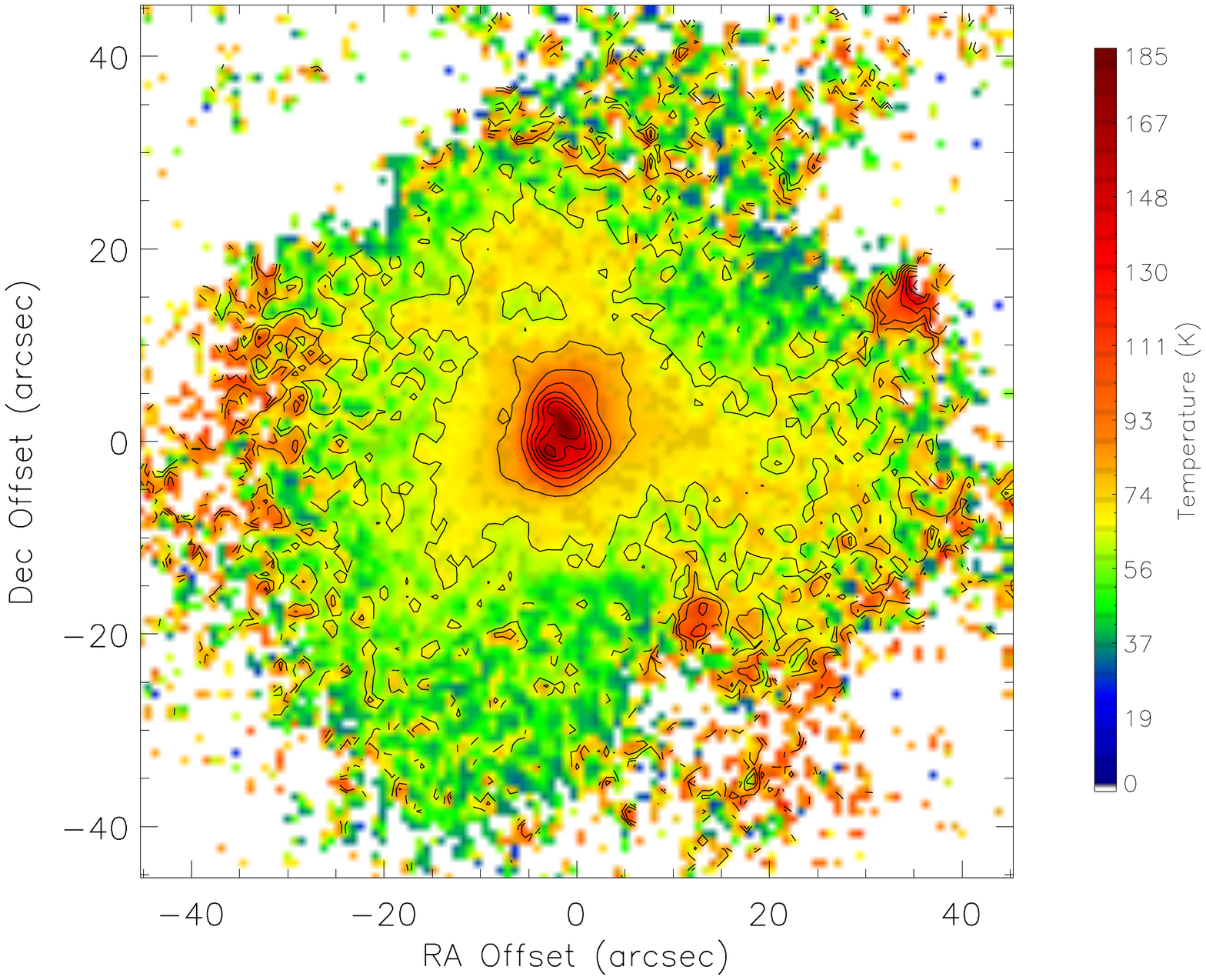} 
}
\end{minipage}
\hspace{-0.25in}
\begin{minipage}{0.5\textwidth}
      \vspace{-1.5in}
         \includegraphics[scale=0.35, angle=90]{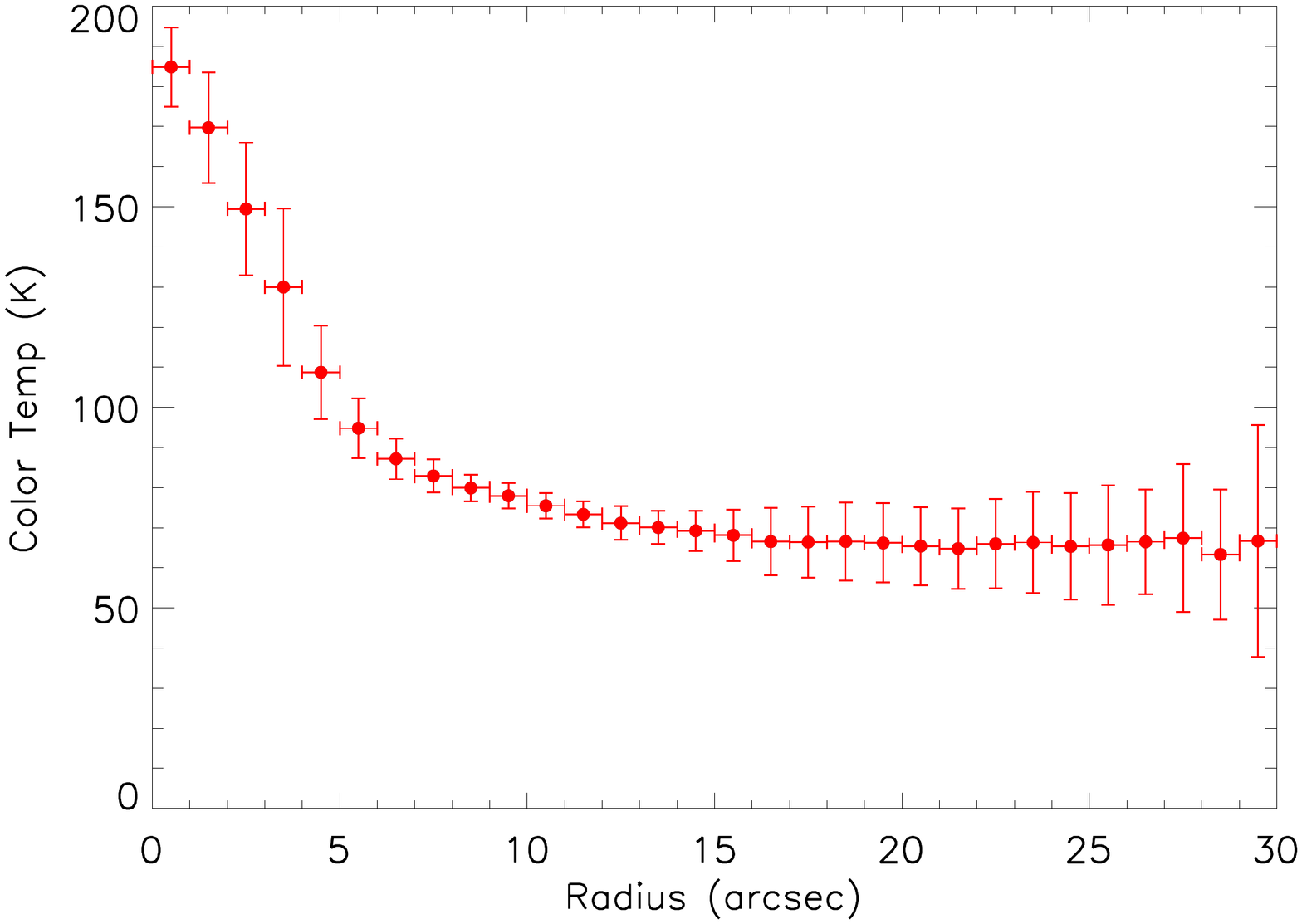}
\end{minipage}
\vspace{-1.75in}
\caption{(left) Dust temperature map derived from the FORCAST data. The contour around the bright
yellow/orange region in the map corresponds to $T \approx 67$ K. (right) Azimuthally averaged radial profile of the dust temperature.
\label{dusttempmap}}
\end{figure*}

In order to determine the properties of the dust in the southwestern arc and northern loop, we 
estimated the dust temperature at each pixel in the 19.7, 31.5, and 37.1 $\mu$m
combined image by fitting the emission with a model given by 
\begin{equation}
F_{\nu} =\Omega \cdot S_{\nu}(T_{dust})  \cdot [1 - \exp(-\tau_{\nu})]
\end{equation}
where $S_{\nu}(T)$ is the blackbody function, $T_{dust}$ is the temperature of
the dust, and $\tau_{\nu}$ is the dust opacity \citep[see e.g.][]{Shuping12}.
We adopted the dust opacity law of \citet{McClure09}. Note that the full-width at half maximum (FWHM) of the point
spread function (PSF) in the FORCAST images does not vary substantially across
this wavelength range. The resulting dust map and radial profile are presented
in Fig. \ref{dusttempmap}. The temperature map indicates that the dust has a relatively uniform
temperature of $T_{dust} \sim 67$K and a low optical depth ($\tau_{\nu} << 1$)
at all of the FORCAST wavelengths.

\subsection{System Inclination}

\begin{figure*}
\begin{minipage}[t]{4.5in}
\hspace{0.5in}
{\includegraphics[scale=0.55]{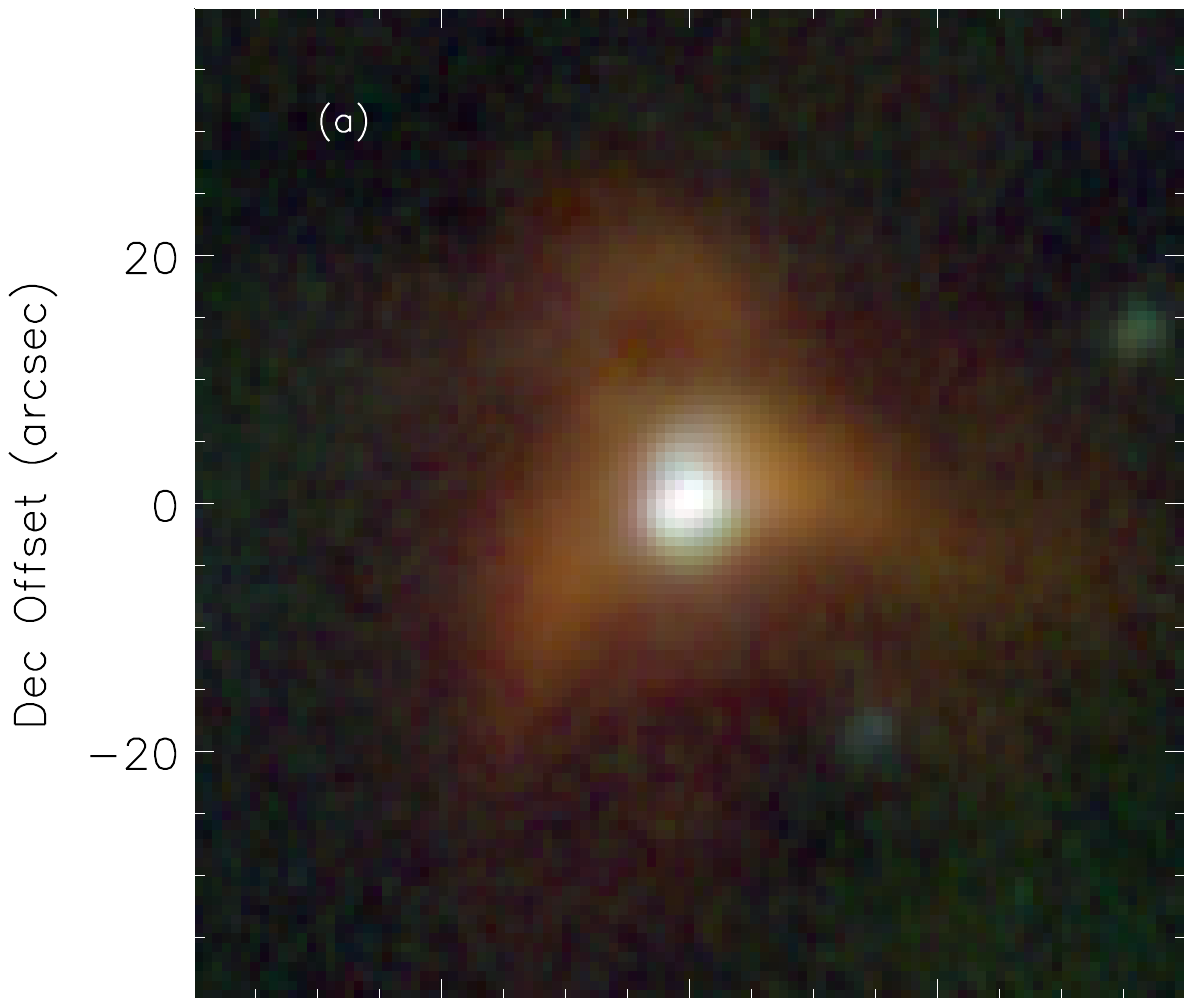} 
}
   \begin{minipage}[t]{4.5in}
      \vspace{-4.01in}
      \hspace{0.5in}
      \includegraphics[scale=0.55]{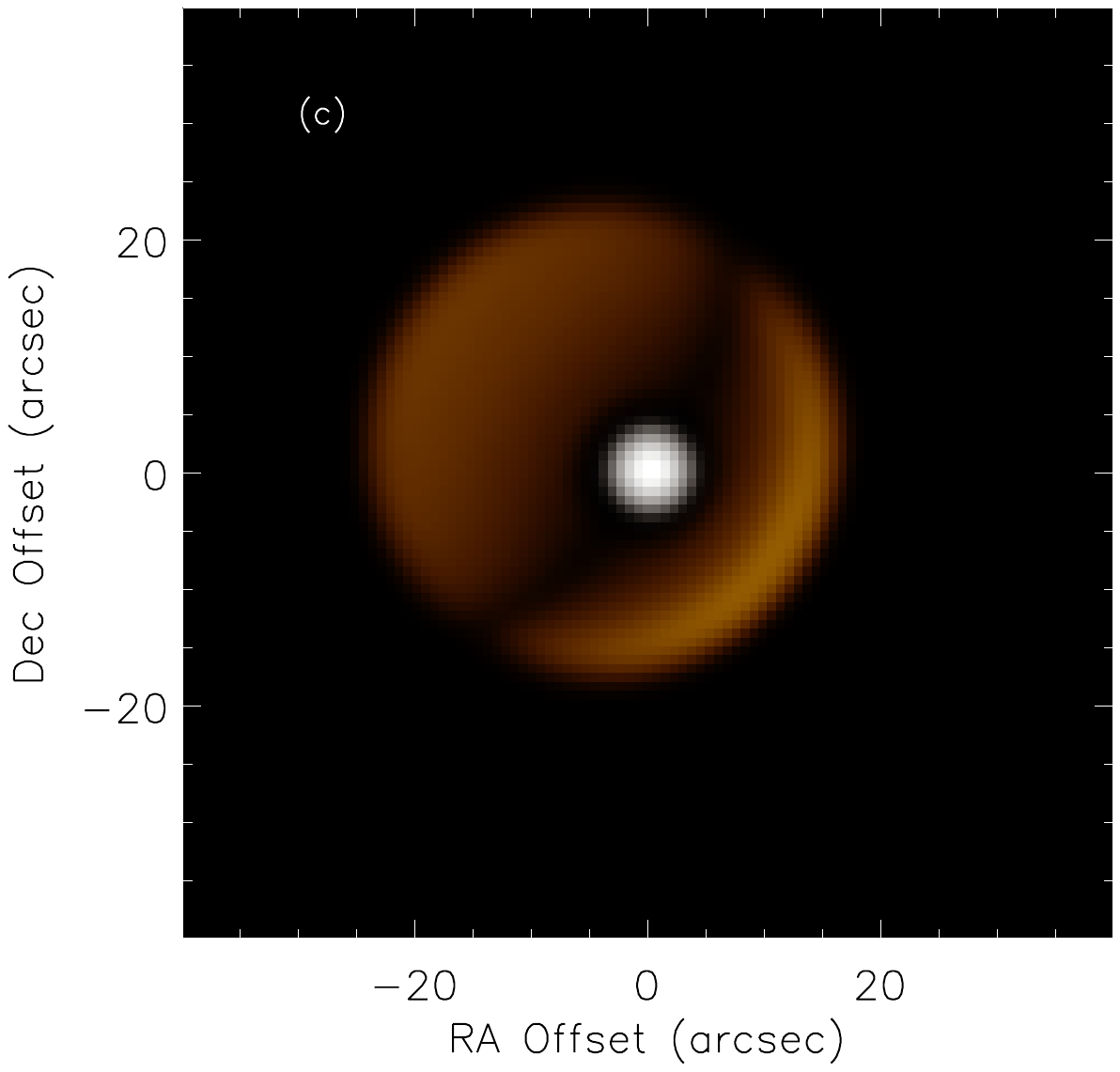}
   \end{minipage}
\end{minipage}
\hspace{-1.905in}
 \begin{minipage}[t]{4in}
      \vspace{-6.05in}
        \includegraphics[scale=0.55]{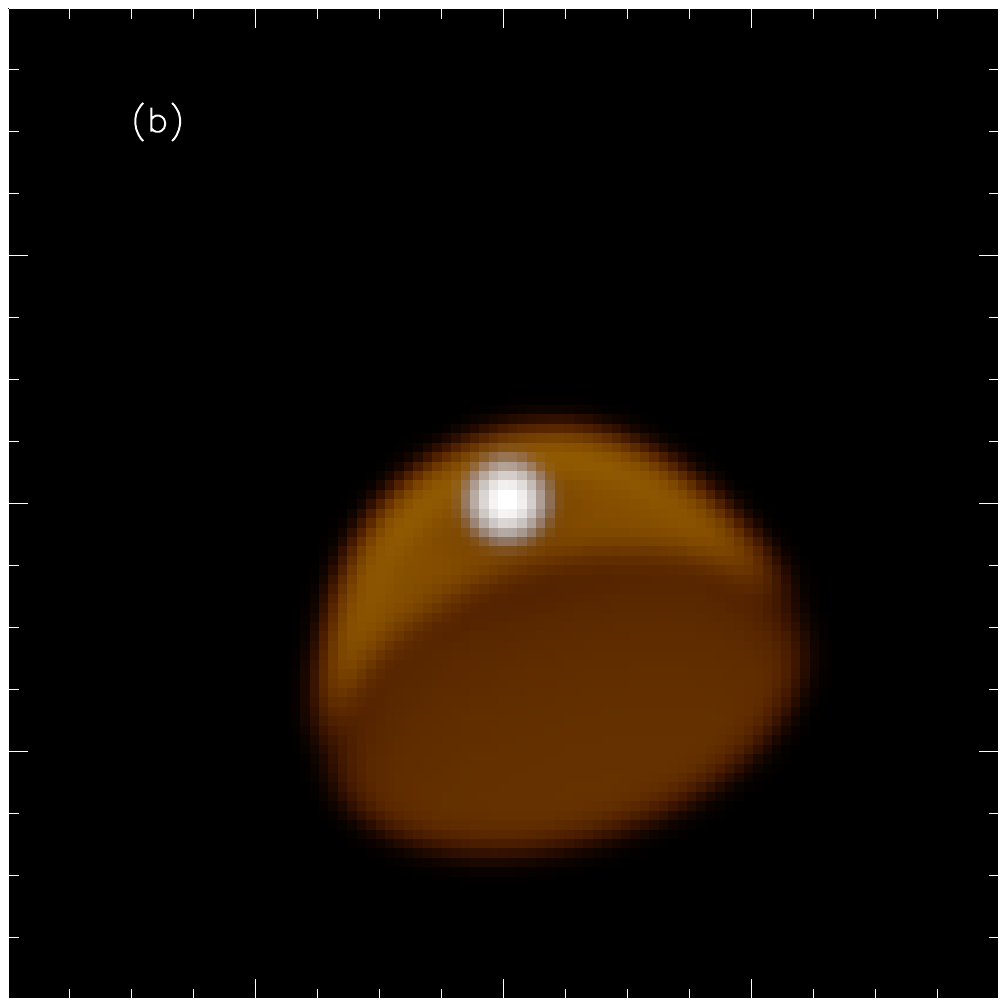}
       \begin{minipage}[t]{4in}
           \vspace{-4.01in}
           \includegraphics[scale=0.55]{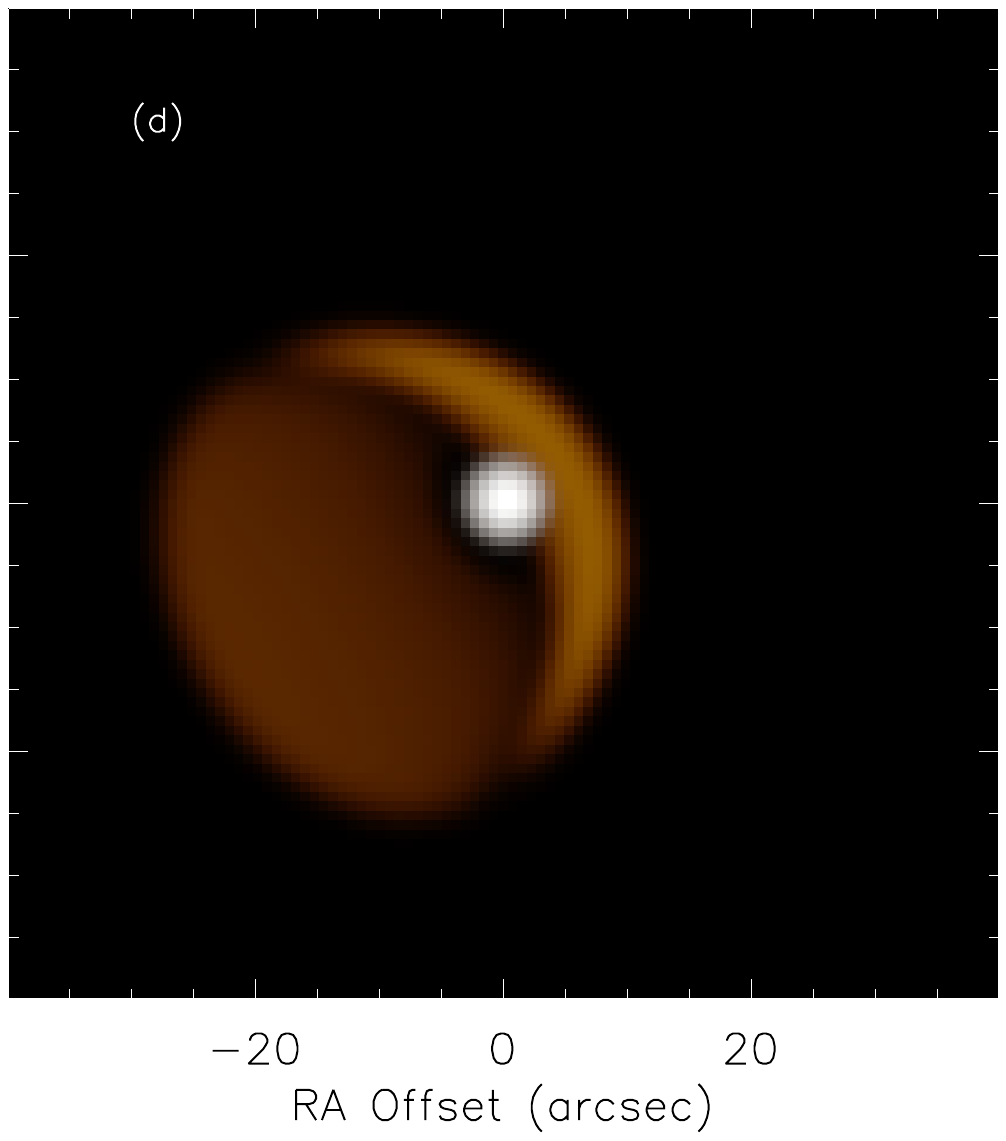}
       \end{minipage}
\end{minipage} 
\vspace{-2.5in}
\caption{(a) Center of the FORCAST 3-color image on an asinh intensity scale;
(b) Simulated image for best fit parameters found in this work (PA $= 197$\degr and $i = 55$\degr) on the same intensity scale;
(c) Simulated image generated for PA $\sim 56$\degr and $i \sim 15$\degr, as found by Malbet et al. (2007);
(d) Simulated image generated for PA $\sim 120$\degr and $i \sim 40$\degr, as  found by  Acke et al. (2008).}
\label{sim_FORCAST_img}
\end{figure*}

To constrain the inclination of the system, we constructed a simple geometrical
model of the outflow structure and synthesized MIR images which could be compared to the
FORCAST data. For simplicity, we limited ourselves to modeling only the southwestern arc , for which the FORCAST
images provide sufficient data to constrain any models. 
We assumed that the dust emission traced the surface of a conical
outflow with a shape given by $r(z) \sim z^\alpha$. The dust emission surface was
assumed to be physically and optically thin, with a uniform thickness, and isothermal with a temperature of 67 K; 
these assumptions are justified by our results described above. In order to simulate the wavelength
dependence of the dust emission, we assumed a wavelength dependence of the dust emissivity of
$Q_{abs} \sim \lambda^{-1.5}$. The dust emission spectrum was scaled to roughly match the 
observed surface brightness in the FORCAST images. The central source was added to the image, with
an intensity at each wavelength equal to that seen in the FORCAST images. The simulated images
were scaled to match the observed physical size of the FORCAST images. 
We rotated the structure in three dimensions, convolved the
resulting image with a Gaussian with a FWHM equivalent to that of the FORCAST
PSF (FWHM $\sim 3''$), and resampled it to match the FORCAST pixel size. 
The free parameters of the model were the power law
index of the conical shape, $\alpha$, the position angle (PA) on the sky, and
the inclination angle, $i$. We generated $\sim 800$ models with a range of parameter values
($\alpha$ was varied between 0.1 and 0.6, $i$ between 35 and $60$\degr, and PA between 
185 and $210$\degr) and attempted to match the resulting image to the observations. 
The best-fit parameters, yielding an image most similar to that of the data, were found to be 
$\alpha = 0.3\pm 0.15$, PA $= 197\pm 10$\degr, and $i = 55\pm 5$\degr. 
Given the limited length of the arc seen in the FORCAST images and the relatively low signal-to-noise 
ratio and contrast of the emission compared to the background, the uncertainties on $\alpha$ and 
the PA are fairly large and a wide range of values provided reasonable fits to the data. However, the PA we find 
agrees well with that determined by \citet{Sallum21} from interferometric imaging
with the LBT. Furthermore, the 
inclination of the conical structure is quite well-constrained. The synthetic image produced by the 
model for these parameters shown in Fig.\ \ref{sim_FORCAST_img} is remarkably close to the 
observations. We then used the same model to generate synthetic images using the parameters 
derived from the analyses of the NIR interferometric data (PA $\sim 56$\degr, and $i \sim 15$\degr from 
\citealt{Malbet07}; PA $\sim 120$\degr, and $i \sim 40$\degr from \citealt{Acke08}), with $\alpha \sim 0.3$. 
The resulting synthetic images are shown in Fig.\ref{sim_FORCAST_img}, which demonstrates that 
these parameters cannot reproduce the structure seen in the FORCAST images.

\subsection{Outflow Parameters}

The IRTF spectrum reveals strong emission lines from the H Paschen, Brackett, and Pfund series, as well as O I lines, atop a strong,
red continuum. The He {\small I} 1.083 $\mu$m line can be seen and exhibits a clear P Cygni profile, with two absorption dips. Weak CO 2-0 bandhead emission can be 
seen at 2.29 $\mu$m. Continuum discontinuities associated with the three H line series are quite prominent and are clearly in emission. 
There is no evidence of any absorption lines associated with the stellar photosphere, which indicates that the near-infrared continuum at 
wavelengths at least as short as 0.8 $\mu$m is dominated by a combination of dust, bound-free, and possibly free-free continuum emission. 

\begin{figure}
\hspace{1.0in}
\includegraphics[scale=0.45,angle=90]{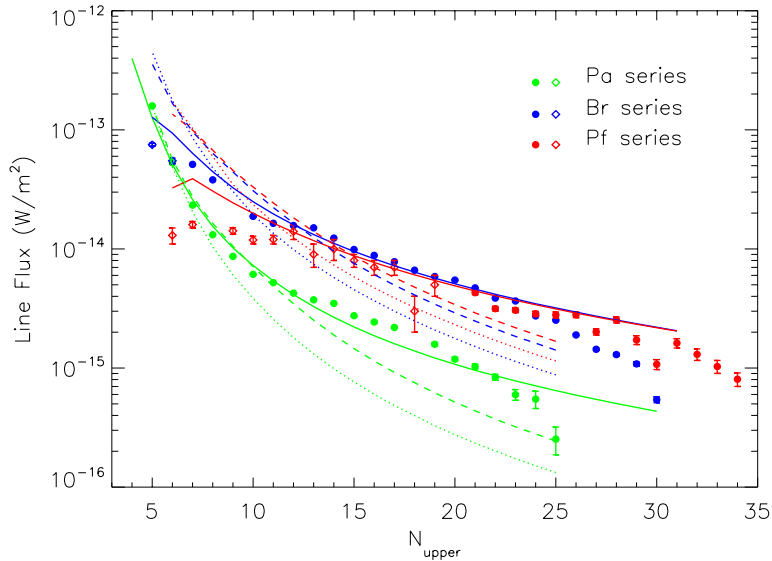}
\vspace{-0.25in}
\figcaption[]{Comparison between observed line fluxes and those predicted by the wind model (solid lines) as a function of the upper level number for the
transition $n_{\rm upp}$. Solid symbols denote measurements from the IRTF/SpeX data, while open symbols are from \citet{Benedettini01}. Dotted lines
are Case B values for $T = 10000$ K, $n_e = 1.0 \times 10^2$ cm$^{-3}$, reddened by $A_V = 8.1$ mag and scaled to match the observed Pa $\beta$ flux,
while the dashed lines are similarly Case B values for $n_e = 1.0 \times 10^{10}$ cm$^{-3}$.
\label{linefluxes}}
\end{figure}

The large number of H lines in the spectrum provides a wealth of data that can be used to infer the
properties of the system. We detect at least 57 H lines in the Paschen (down to the Pa 25 line), Brackett (to Br 30), and Pfund (to Pf 34)
series, along with lines of He {\small I} and O {\small I}, with a signal-to-noise ratio considerably higher than in the spectra 
of \citet{Benedettini01}. 
As shown in Fig \ref{linefluxes}, the line fluxes and flux ratios estimated from the standard Case B assumption \citep{StoreyHummer95}
do not provide a good fit to the entire set of observed values for any values of temperature and electron density \citep[see also][]{Drew97,Benedettini98}. 
Following \citet{Nisini95} and \citet{Benedettini01},
we therefore constructed wind models, using the Sobolev approximation
\citep{Sobolev60} and the formalism developed by \citet{Castor70}, to predict
the strengths of hydrogen emission lines under the assumption of a two-level
atom \citep[see e.g.,][chap.\ 9]{Peraiah02}. The models assume that the wind is isothermal and fully ionized and that LTE conditions apply.
An extensive discussion of the applicability and limitations of these assumptions for wind models of HAeBe stars is given by \citet{Nisini95}.
The assumed velocity profile is given by
\begin{equation}
v(r) = v_i + (v_{max} - v_i) \Bigg (1- \bigg (\frac{r_i}{r} \bigg )^\alpha \Bigg )
\end{equation}
where $v_i$ is the velocity at the base of the wind, located at $r_i$, and $v_{max}$ is the terminal velocity of the wind. This
expression for the velocity profile is commonly used in the analysis of hot star winds \citep[see][]{Groenewegen89}.
The mass loss rate $\dot{M}$ for the wind is then given by
\begin{equation}
\dot{M} = 4\pi r^2 m_p n_e(r) v(r)~ .
\label{mdot}
\end{equation}
We did not incorporate variations with wind velocity and mass flux with stellar latitude \citep{Malbet07}.
In these models, the flux density at a frequency $\nu$ arising from a line transition at frequency $\nu_0$, 
with statistical weight of the lower level $g_l = 2 l^2$, oscillator strength $f_{lu}$, and energy of the lower 
level of $E_l = 13.6/l^2$, is given by 
\begin{equation}
F_\nu = \frac{2\pi S_{\nu}(T)}{D^2} \int_{r_i}^{r_o} (1 + \mu^2 (\nu, r) \sigma(r)) [1 - \exp(-\tau(\mu,\nu,r))] r dr
\end{equation}
where $S_{\nu}(T)$ is the Planck function at temperature $T$, which we have assumed is constant throughout the wind. 
Here $\mu$ is the angle between the radial direction and the line of sight and is given by
\begin{equation}
\mu = \frac{c}{v(r)}\frac{\nu - \nu_0}{\nu_0}, 
\end{equation}
while
\begin{equation}
\sigma(r) = \frac{r}{v(r)}\frac{dv}{dr} - 1
\end{equation}
and 
\begin{equation}
\tau (\mu,\nu,r) = \frac{\tau_0(r)}{1+ \mu^2 (\nu, r) \sigma(r)}
\end{equation}
where
\begin{equation}
\begin{split}
\tau_0(r) = \frac{\pi e^2 n_e^2 r}{2 m_e v(r)\nu_0} \Bigg ( \frac{h^2}{2\pi m_e kT}\Bigg )^{1.5}  g_l f_{lu} \exp(E_l/kT) [1 - \exp(-h\nu_0/kT)]
\end{split}
\end{equation}

To keep the number of free parameters relatively small, we assumed $D=418$
pc, $T=10000$ K, $v_i = 20$ km s$^{-1}$ (in agreement with \citealp{Nisini95} and
\citealp{Benedettini01}) and $\alpha = 1$ (in agreement with the
description of the winds for massive stars; \citealp[e.\ g.][]{Puls96}). Although we explored
models with a variety of values for $r_i$, we decided to adopt $r_i =
R_\star = 6 R_\odot$. This left us with $\dot{M}$, $v_{max}$, and $\rho = r_o/r_i$
as the free parameters. Based on the observed line widths (mean FWHM $\sim 200$
km s$^{-1}$) and the inherent resolution of the instrument ($R \sim 2000$), we
limited $v_{max}$ to be less than or equal to 200 km s$^{-1}$.

\begin{deluxetable*}{lcl}
\tabletypesize{\scriptsize}
\tablecolumns{3}
\tablenum{7}
\tablewidth{8in} 
\tablecaption{Adopted Values for Wind Model Parameters 
\label{tbl-modpars}} 
\tablehead{
\colhead{Parameter}   &  \colhead{Values} & \colhead{Units}\\
}
\startdata
$\alpha$ & 0.0, 1.0 \\
$v_{max}$ & 50, 75, 100, 150, 200 & km s$^{-1}$ \\
$\dot{M}$ & $5\times 10^{-8}$, $(1,3,5,5.5,6,6.5,7,7.5,8,8.5,9,9.5)\times 10^{-7}$, $(1,1.1,2,3)\times 10^{-6}$ & $M_{\odot}$ yr$^{-1}$ \\
$\rho$ & 2, 5, 10, 15, 20, 25, 30, 35, 40, 45, 50, 75, 100, 125, 150, 200 & \\
$A_V$ & 4 - 12 in steps of 0.1 & mag\\
\enddata
\end{deluxetable*}

\begin{figure}[t]
\includegraphics[scale=0.675, angle=90, trim=0in 2in 0in 0in]{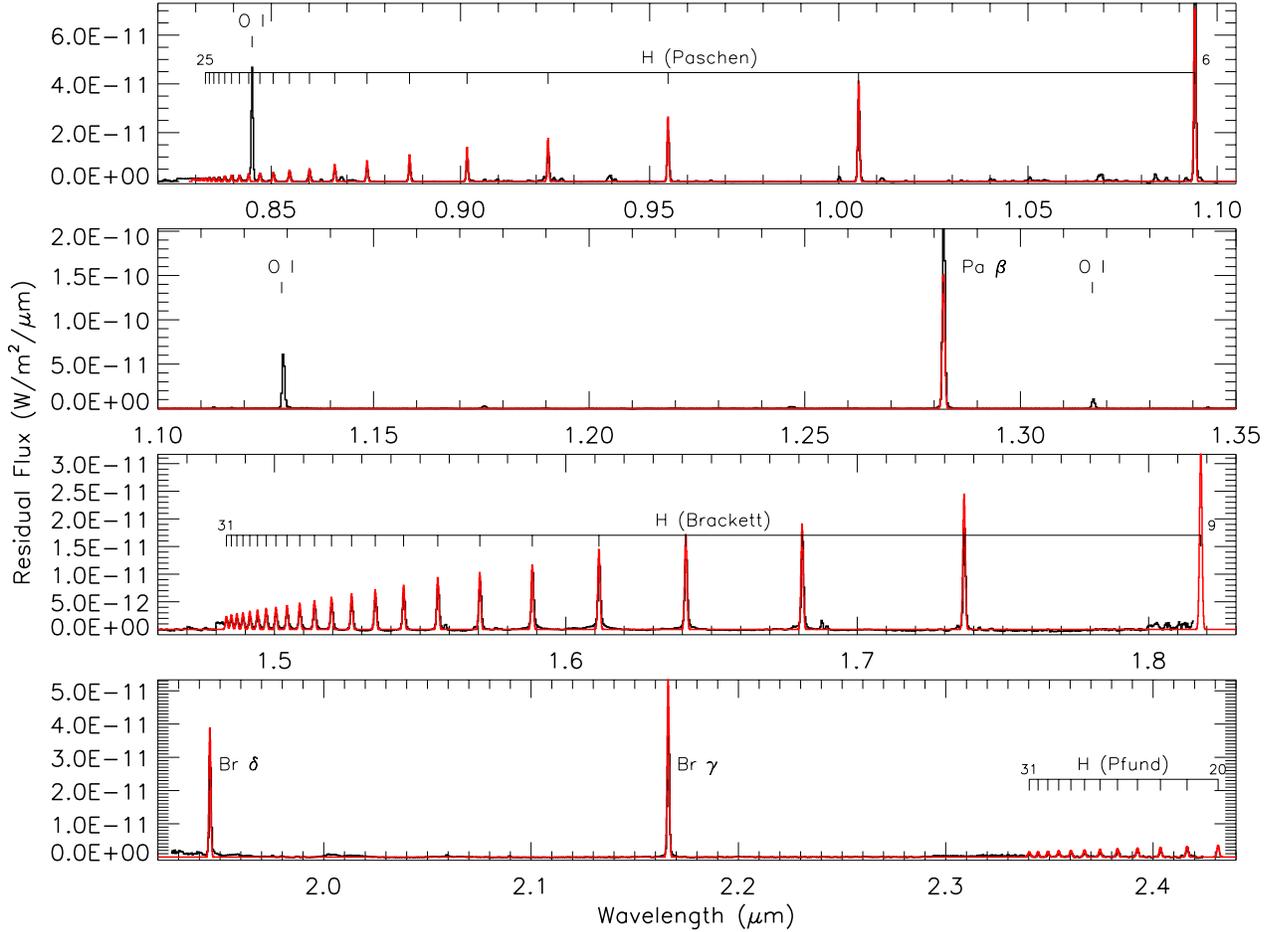}
\vspace{-0.25in}
\figcaption[]{Comparison between the continuum-subtracted IRTF/SpeX 
spectrum and the best fitting wind model (red). Only H lines have been synthesized in the model.
\label{speccomp}}
\end{figure}

\begin{figure}
\hspace{1.2in}
\includegraphics[scale=0.5, angle=0, trim= 0in 4in 0in 0in]{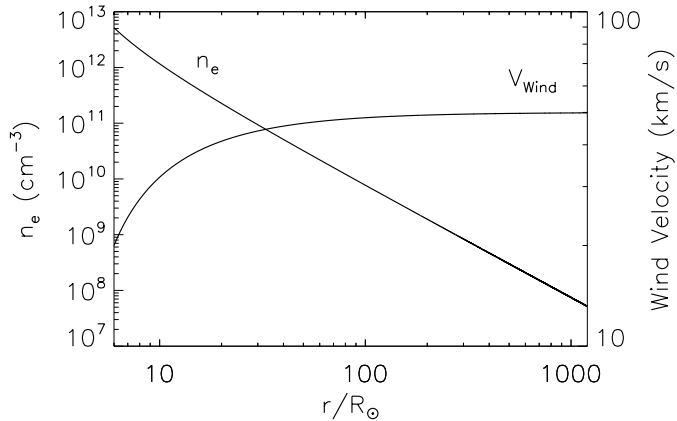}
\figcaption[]{Radial profile of the electron density (left hand scale) and wind velocity (right hand scale) in H line emitting region
for our best fit parameters and the adopted wind model.
\label{neden}}
\end{figure}

We generated over $1300$ models with a range of parameter values centered on those found by \citep{Benedettini01};
the range of values adopted for the various parameters are given in Table \ref{tbl-modpars}.
For each model we computed the fluxes of the H lines of the Paschen,
Brackett, and Pfund series up to at least $n_{\rm upp}=30$. We then reddened the line
fluxes using the \citet{Cardelli89} extinction model and $A_V$ values between 4
and 12 mag in steps of 0.1 mag. This resulted in over $10^5$ reddened models. We compared these reddened 
models with the observed line fluxes and determined the best fitting model by computing $\chi^2$. We found
that the best fit to the observed line strengths were provided by the model with
$\rho \ge 200$, $v_{max} = 50$ km s$^{-1}$, $\dot{M} = 6.0 \times 10^{-7} M_\odot
~\mathrm{yr}^{-1}$ and $A_V = 8.1$ mag. A comparison between the observed line
fluxes and those of the best fit model are shown in Fig \ref{linefluxes}. The
model spectrum, convolved to the SpeX resolution ($R\approx 2000$), is plotted along with
our continuum-subtracted SpeX spectrum in Fig. \ref{speccomp}. The latter figure
indicates that the model provides a very good match for most of the observed
lines, although it slightly underestimates the strength of the Pa $\beta$ line (in contrast to the results of \citealt{Nisini95}) and 
overestimates the flux of the highest and lowest Br transitions. All of the lines, even those arising 
from high upper levels, are found to be optically thick. The radial profiles of the 
wind velocity and the electron density in the wind, computed with our best fit 
parameters are shown in Figure \ref{neden}. Based on the $\chi^2$ fits, we estimate the
the allowed ranges on the reddening and mass-loss values to be $A_V = 8.1 \pm^{2.5}_{1.5}$ 
mag and $\dot{M} = 6.0 \pm ^{3.7}_{1.7} \times 10^{-7} M_\odot~\mathrm{yr}^{-1}$. Due to 
our model sampling, useful constraints cannot be derived for $v_{max}$ although the derived
value is consistent with the observed widths of the strongest emission lines in the SpeX spectrum. 
Constraints on $\rho$ are also difficult to obtain, as almost all values of $\rho > 25$ yielded acceptable fits.
Models with $\rho$ values up to as large as 2000 were
generated and slightly better fits to the observed spectrum were found as $\rho$ increased, although
the improvement was minuscule and almost entirely due to a tiny increase in the predicted Pa $\beta$ line flux. On
the other hand, we can estimate the radius of the line emitting region by assuming that this radius represents
the surface at which $\tau = 1$, and the line emission is given by a blackbody with temperature $T$ \citep[see][]{Nisini04}. 
In this case,
\begin{equation}
r_o = \Bigl[\frac{c L_{\rm line}}{4 \pi^2 B_{\lambda}(\lambda, T) \lambda \Delta v}\Bigr]^{0.5} ~~~,
\end{equation}
where $L_{line}$ is the luminosity in the line, and $B_{\lambda}(\lambda, T)$ is the Planck function. 
For both Pa $\beta$ and Br $\gamma$, with $T = 10000$ K, and $\Delta v = 50$ \kms\ we find $r_o \lesssim 1$ au, or $\rho \lesssim 30$.
The best fit value of $\rho = 200$ corresponds to $r_o = 5.6$ au.

Our estimate of the reddening is in good agreement with the values found in the literature \citep[e.g.][]{Nisini95,Drew97,Benedettini01,Fairlamb15,Vioque18,Ubeira20,Wichittanakom20}.
Furthermore, the value of $v_{max}$ we find is consistent with that found by \citet{Drew97} for the Br$\alpha$ line, although it is significantly smaller
than the value adopted by \citet{Nisini95} based on the width of the H $\alpha$ line in the spectrum shown by \citet{Finkenzeller84}. 
Their value of 380 \kms is not supported by the observed widths of the emission lines in our spectrum.
The mass-loss rate derived from our best fit model is about 30\% of the value found by \citet{Benedettini01} and
about half that given by \citet{Nisini95} (both for a distance of 450 pc), while our reddening 
estimate is only slightly larger. This is perhaps unsurprising given that variations in the 
widths and strengths of the strongest H lines in MWC 297 have been reported in the literature \citep[see e.g.][]{Drew97,Benedettini01,Eisner15}. 
Our measured line strength for the Pa $\beta$ line is about half that given by \citet{Nisini95}  while our
Br $\gamma$ line flux is about 16\% larger. We note that our mass-loss rate is about a factor 5 smaller than that 
derived by \citet{Jinliang97} using non-LTE wind models and the line fluxes reported by \citet{Nisini95}.

The tendency of the wind models
to overestimate the strengths of the weakest transitions (i.e., those with the highest upper levels, $n_{\rm upp}$) was 
before noted by \citet{Nisini04}, who suggested that this may indicate that the wind velocity
law is incorrect. While a much steeper acceleration, occurring over a much smaller radius, may
possibly yield better fits, such a velocity law seems unphysical and is inconsistent with the 
description of winds from massive stars. While a full investigation of the wind velocity law is beyond the scope of this paper, 
we did generate a set of models with $\alpha = 0$, and a smaller set with $\alpha=10$, as extreme examples of a wind that is at, or close to, the terminal velocity 
when it leaves the stellar surface. We did not find that these models produced significantly better fits to the data than those
with $\alpha = 1.0$; better fits to the strengths of the weaker lines in each series were obtained at the expense of poorer
fits to the lines with intermediate values of $n_{\rm upp}$. Significantly, we also found that the parameters of the best fit model to the observed line fluxes 
were very similar to those we found above ($\dot{M} = 8.5 \times 10^{-7} M_\odot~\mathrm{yr}^{-1}$, $A_V = 8.2$ mag,
$v_{max} = 75$ \kms, and $\rho = 200$).

The H line strengths provide an estimate of the ionizing luminosity. If the lines were optically thin, we could use the predicted emissivities tabulated
by \citet{StoreyHummer95} for Case B conditions to estimate the ionizing luminosity producing the H lines. Adopting a gas temperature of $10^4$ K and an 
electron density of $10^9$ cm$^{-3}$, using the observed strengths of 
the Br $\gamma$ lines, correcting for extinction corresponding to $A_V = 8.1$ mag, and adopting a distance of 
418 pc, we find the ionizing photon flux to be $Q_0 \sim 2 \times 10^{47}$ photons s$^{-1}$. Of course, our models 
indicate that none of the H lines are optically thin, and therefore the Case B estimate of $Q_0$ is not strictly applicable,
but it can be considered a lower limit to the true ionizing flux in the system. We can also use the equation from \citet{Felli81} to
estimate the ionization rate for an accelerating wind for a given mass-loss rate. For the mass-loss rate we find from our models, 
and the velocity law we have adopted, we find $Q_0 = 2.2 \times 10^{48}$ photons s$^{-1}$. 
Using a Kurucz model with
a temperature of 24000 K, a $\log g$ of 4.0, and a radius of $6 R_{\odot}$, parameters typical of a B1.5 V star \citep{BohmVitense81,SchmidtKaler82}, 
we estimate the ionizing photon flux generated by the star to be $Q_0 = 3.8 \times 10^{45}$ photons s$^{-1}$, a factor of at least $50$ smaller than 
the estimated value. A similar, but slightly smaller, value of $2.9 \times 10^{45}$ photons s$^{-1}$  is found using the B star models of 
\citet{LanzHubeny07}. With our assumed velocity law, the maximum mass-loss rate for which a B1.5V star can completely ionize its wind is only
$\sim 2.5  \times 10^{-8} M_\odot~\mathrm{yr}^{-1}$ (see \citealt{Felli81}),  clearly far below the value we estimate from the emission lines in our SpeX data.
This result suggests that either the wind is completely ionized within only a tiny region beyond the stellar photosphere (which seems incompatible 
with our model results) or nearly all of the ionizing luminosity is generated by the accretion process (e.g., by shocks, as the accreting material impacts 
the stellar surface), not the star itself. 

The Br $\gamma$ luminosity is often used as a measure of the accretion rate of a young stellar object. In the case of MWC 297, our
dereddened luminosity is $L({\rm Br} \gamma)/L_{\odot} \approx 0.67$.  Using the correlation between $L({\rm Br}\gamma)$ and the accretion luminosity $L_{acc}$
given by \citet{Fairlamb17}, we find $L_{acc} \approx 1.7 \times 10^4 L_{\odot}$, which in turn corresponds to a mass accretion rate of 
${\dot M}_{acc} \approx 3.3 \times 10^{-4} M_{\odot}$ yr$^{-1}$, or about 540 times larger than the mass-loss rate we derive from our wind models. If we use 
the observed Pa $\beta$ line strength, we estimate 
$L_{acc} \approx 3.2 \times 10^4 L_{\odot}$ and ${\dot M}_{acc} \approx 6.2 \times 10^{-4} M_{\odot}$ yr$^{-1}$, nearly a factor of two larger. Such large
accretion rates accord with our observation of an optical and NIR continuum dominated by emission lines and bound-free continuum, with the Paschen,
Brackett, and Pfund jumps all in emission. The accretion rate we derive from the Br $\gamma$ luminosity is in good agreement with the value estimated 
by \citet{Wichittanakom20} from the H $\alpha$ line strength. 
However, it is a factor of $\sim 9$ times larger than that given by \citet{GuzmanDiaz21} and $\sim 8$ times larger than that derived by \citet{Fairlamb17} (after accounting for
the different distance adopted). 

\subsection{SED}

To generate the spectral flux and spectral energy density (SED) of MWC 297, we began by compiling and carefully examining the data in the voluminous literature
on this object, using the table in Appendix A.9 of \citet{Alonso-Albi09} as an initial guide.
In addition to our SOFIA/FORCAST, 
Herschel/PACS, and Herschel/SPIRE measurements, we included photometry from a number of individual measurements reported in the literature as well as 
sky surveys, including Pan-STARRS \citep{Chambers16}, GAIA \citep{Riello21}, APASS \citep{Henden15}, SkyMapper \citep{Wolf18}, TASS \citep{Richmond00}, 
FONAC-S \citep{Yuldoshev17}, DENIS \citep{Kimeswenger04}, 2MASS, IRAS, and  MSX. Filter wavelengths were adopted from the various references or,
in the case of the Johnson/Cousins system, computed using the response functions given by \citet{Mann15}. Similarly, magnitudes were converted to fluxes in
Jy using the zero points in the references or from \citet{Mann15}.
We then added our own submm/radio measurements, as well as those of \citet{Skinner93}, \citet{Mannings94}, \citet{Henning98}, \citet{Manoj07}, and 
\citet{Alonso-Albi09}. If multiple measurements at the same wavelength were
available, we computed a weighted average. The full list of adopted fluxes is given in Table\ref{litfluxes}.

\begin{figure}
\begin{minipage}[t]{8.5in}
\hspace{1.5in}
{\includegraphics[scale=0.35,angle=90,trim= 1in 2in 0in 0in]{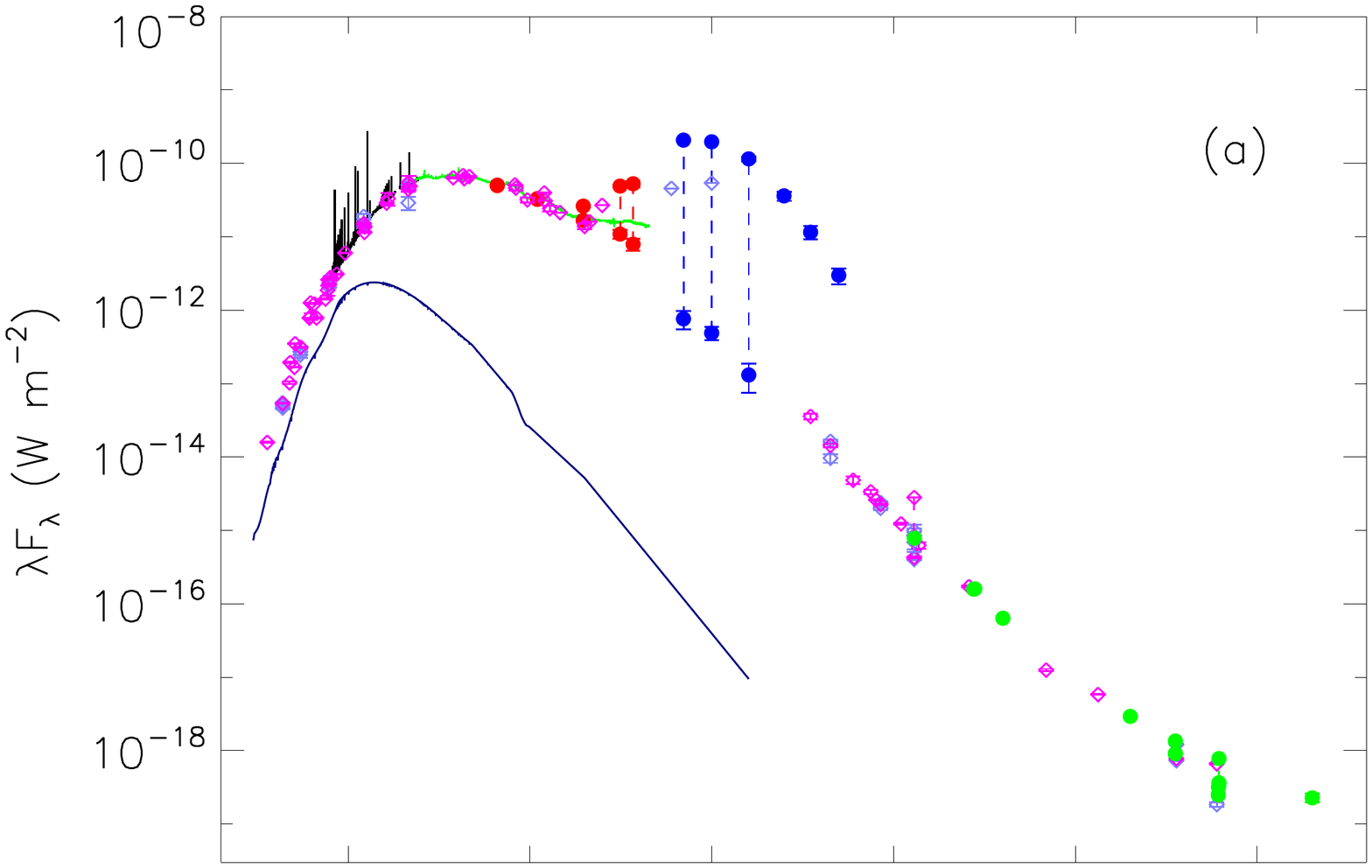}
}
\end{minipage}
\begin{minipage}[t]{4.5in}
      \hspace{1.51in}
      \includegraphics[scale=0.35, angle=90,trim= 1in 0in 2.0in 0in]{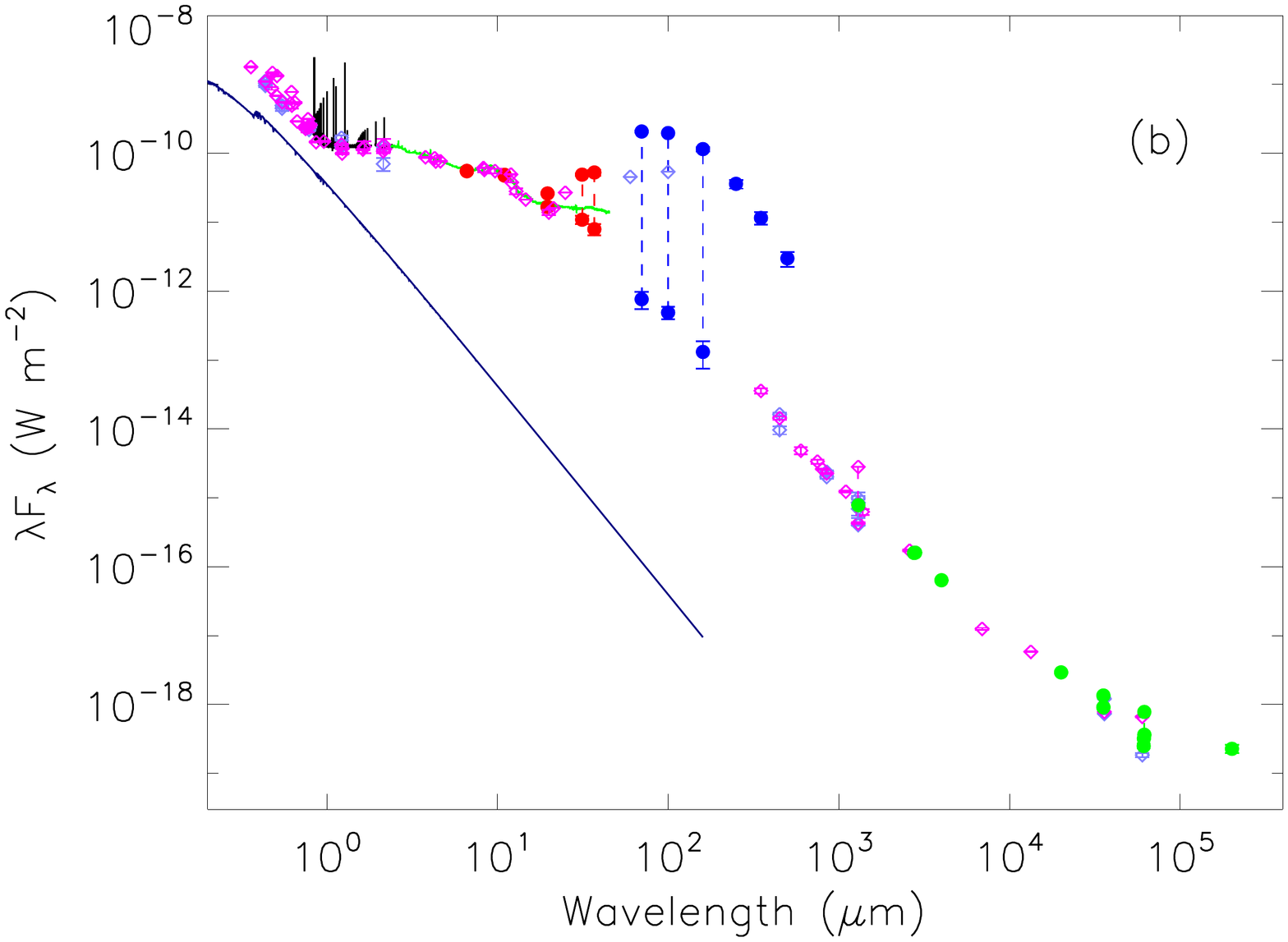}
\end{minipage}
\vspace{0.25in}
\figcaption[]{\label{fig-sed_full}(a) Observed  and (b) dereddened SED of MWC\,297. Red points are the values derived from SOFIA/FORCAST, while
blue points are from Herschel/PACS and SPIRE and green points are from newly measured submm/radio observations. Pink
diamonds are values taken from the literature. Blue diamonds are literature values that were not considered in the fit or
were combined to generate values shown in pink. Vertical dashed lines connect values for the core and the extended region. Solid
black line is the IRTF/SpeX spectrum and solid green line is the ISO spectrum. Solid dark blue line is a Kurucz model for a
B1.5 V star, reddened by $A_V = 8.1$ mag in (a). }
\end{figure}

In Figure~\ref{fig-sed_full} we present the full spectral energy distribution of MWC\,297 from 0.36 $\mu$m to $6.0$ cm.
We included the archival ISO spectrum
and our SpeX spectrum (scaled by a factor of 1.22) in the plot. Values that were used to generate averages are plotted in light blue. 
Dashed lines connect photometric measurements for the core and and core+extended emission
at the same wavelength. This figure demonstrates the remarkable agreement between the
flux levels of the various data sets, at least in the NIR and beyond, including the SOFIA photometry. At wavelengths longwards of
20 $\mu$m the ISO spectrum deviates from the other measurements
because the rectangular ISO aperture captured only a fraction of the total flux in the extended envelope. 

 The flux for a Kurucz model with $T_{eff} = 24000$K, $\log g = 4.0$, and $R = 6 R_\odot$, corresponding to parameters for a B1.5V star,
at a distance of 418 pc and reddened by $A_V = 8.1$ mag, is also shown in Figure~\ref{fig-sed_full}. The fluxes for this stellar model are well
below the optical photometry points. A good fit to the optical photometry can be obtained if the reddening is reduced ($A_V \sim 7$ mag) or
the stellar radius is increased ($R \sim 10 R_{\odot}$). Radius values larger than $10 R_{\odot}$ yield model spectra that are incompatible with the blue end of our SpeX spectrum
while models with radii larger than $\sim 7 R_{\odot}$ are incompatible  with the critical rotation velocity (see Section 4.1). 
Irrespective of the specific parameter values, it is clear from both this figure and our
SpeX spectrum (Fig. \ref{fig-SpeX}) that the flux longwards of about 0.8 $\mu$m is dominated by continuum emission that is not from the stellar photosphere,
but must be generated by the accretion process. 

To estimate the total luminosity, we fit a smooth curve to the photometric points shown in Figure~\ref{fig-sed_full} and integrated under the curve
between 0.35 and $6.0$ cm. We find that the core of MWC297 generates 
a luminosity of approximately $740 ~L_\odot$, far below that expected for an early B star ($\sim 1.1 \times 10^4 L_\odot$). Including the flux from the extended 
emission region, under the assumption that this represents re-processed UV emission arising from dust which has been heated by MWC 297, we
find a total luminosity of only $\sim 2300 ~L_\odot$. If MWC297 is an early B star, then clearly the dust that provides the optical reddening is not the 
same dust that is re-processing the UV flux from the star and re-radiating it in the infrared. If, instead, we assume that the reddening is produced by 
dust along the line of sight, but distant from MWC297, as suggested by \citet{Acke08}, then we should correct the optical and near-infrared photometry 
for this absorption before integrating to estimate the
total luminosity. Adopting $A_V = 8.1$ mag, we find that the dereddened core of MWC 297 produces a total luminosity of about $5300~ L_\odot$. Including 
the emission from the extended region yields a total luminosity of about $6900 ~L_\odot$, still somewhat below what one would expect from a B1.5V star.
Of course, we are missing the flux emitted between the UV and the optical $U$ band which could account for some of the difference. For the Kurucz model, the 
luminosity emitted between 2500 and 3500 \AA\ is equal to that emitted longward of 3500 \AA\ and represents 10\% of the bolometric luminosity, while the luminosity emitted 
shortward of 2500 \AA\ constitutes 80\% of the total. For a model with $R = 6 R_\odot$, the luminosity emitted between between 2500 and 3500 \AA\ is $\sim 1070 ~L_{\odot}$.
If we assume that the dereddened luminosity computed from the observed SED represents the combination of
the luminosity emitted $\geq 3500$ \AA\ and all of the luminosity emitted $\leq 2500$ \AA, which has been absorbed by dust and re-emitted, and we account for the missing flux between 2500 and 3500 \AA\, then we estimate a total 
luminosity for MWC 297 of about $7900 L_\odot$, which is slightly below but in reasonable agreement with the luminosity expected from a B1.5V star. Because some of the luminosity must arise from the accretion process, our values represent upper limits to the luminosity of MWC 297 itself.

Analysis of the observed SED also allows us to put upper limits on the reddening and luminosity of the system. For values of the reddening $A_V > 8.5$ mag, the dereddened optical
photometric points deviate substantially from the stellar model spectra at the blue end. Adopting $A_V = 8.5$ mag as the upper limit to the reddening, we find that the core of MWC generates a luminosity of $7540 ~L_\odot$,  or $9130 ~L_\odot$ if the emission from the extended region is included. Adding the missed luminosity between 2500 and 3500 \AA\
then yields an upper limit to the luminosity of MWC 297 of $10200 L_\odot$, in good agreement with that expected from a B1.5V star.
We note that the total luminosity we calculate directly from the SED, even after accounting for reddening and missing flux at unobserved wavelengths, is substantially below the 
values that can be found in the literature \citep[e.g.][]{Fairlamb15,Vioque18,Ubeira20,GuzmanDiaz21}, which are typically derived by choosing a stellar radius such that a $T \approx 24000$ K stellar model
fits the optical photometry. As stated above, and explained below, the stellar radius cannot be significantly
larger than $R = 6 R_\odot$. (A radius of $R = 7 R_\odot$ would increase the luminosities due to the correction for missed flux between 2500 and 3500 \AA\ by less than $400~L_\odot$.)
If we adopt a smaller reddening value, in which case the B1.5V  ($T=24000$ K) stellar model matches the {\it observed} SED in the optical, the
discrepancy between the expected luminosity and the value derived from the {\it dereddened} SED becomes larger. 
\begin{deluxetable*}{cccl}
\tabletypesize{\scriptsize}
\tablecolumns{4}
\tablenum{8}
\tablewidth{8in} 
\tablecaption{Adopted fluxes for MWC\,297 from catalogs and the literature \label{tbl-6}} 
\tablehead{
\colhead{Wavelength}   &  \colhead{Obs. Flux} & \colhead{Uncertainty} & \colhead{Band, Reference} \\
\colhead{($\mu$m)} &  \colhead{(Jy)} & \colhead{(Jy)} &  
}
\startdata
0.3595     &              0.00194   &       0.00004       & U, \citet{DeWinter01} \\
0.4368     &             0.0078   &       0.0001         &  B, weighted average of FONAC-S \citep{Yuldoshev17}, APASS \citep{Henden15}, and \citet{DeWinter01} \\
0.4763      &            0.0163   &       0.0007          & g', APASS \citep{Henden15} \\
0.481        &           0.0311     &     0.0006\tablenotemark{a}          & g, Pan-STARRS  \citep{Chambers16}  \\
0.5075      &            0.0283    &      0.0003          & g, Skymapper \citep{Wolf18} \\
0.51097    &             0.0597   &       0.0006          & $G_{BP}$, Gaia \citep{Riello21} \\
0.5486      &            0.0573     &     0.0005          & V, average of TASS \citep{Richmond00}, APASS \citep{Henden15}, and \citet{DeWinter01} \\
0.6138     &             0.1605     &     0.0016          & r, Skymapper \citep{Wolf18} \\
0.62179   &              0.2616    &     0.0009          & G, Gaia  \citep{Riello21}\\
0.6247     &             0.1723     &     0.0173          & r', APASS \citep{Henden15} \\
0.6523     &            0.2649      &     0.0034          & R$_{C}$, \citet{DeWinter01} \\
0.671       &            0.1759      &    0.0032          & r, Pan-STARRS \citep{Chambers16}\\
0.752       &            0.3576      &    0.0066          & i, Pan-STARRS \citep{Chambers16}\\
0.7718     &             0.4808     &     0.0744          & i', APASS as reported by \citet{Zacharias15} \\
0.7769     &             0.6731     &     0.0056          & $G_{RP}$ Gaia \citep{Riello21} \\
0.788       &            0.5923      &    0.0098          & I, DENIS \\
0.8007     &             0.7470     &     0.0146          & I$_{C}$, weighted average of TASS \citep{Richmond00} and \citet{DeWinter01} \\
0.866       &            0.8991      &    0.0166          & z, Pan-STARRS \citep{Chambers16}\\
0.962       &            1.9282      &    0.0355          & y, Pan-STARRS \citep{Chambers16}\\
1.221\tablenotemark{b}       &            6.1166       &   0.2225          & J, DENIS, average of DENIS catalog and value given by \citet{Kimeswenger04} \\
1.23         &           4.72          &  0.13            & J, \citet{Berrilli92} \\
1.235       &            5.64\tablenotemark{c}         &   0.10          & J, 2MASS \\
1.63         &           15.7          &  0.29            & H, \citet{Berrilli92} \\
1.662       &            18.3         &   3.6             & H, 2MASS \\
2.144       &            31.8076    &     2.421       & K, DENIS, average of DENIS catalog and value given by \citet{Kimeswenger04} \\
2.159       &            39.9         &   8.7             & K$_s$, 2MASS \\
2.19         &           35.6          &  0.66            & K, \citet{Berrilli92}  \\
3.79         &           80.56        &   1.48            & L, \citet{Berrilli92} \\
4.294\tablenotemark{d}       &            97.89       &    0.01            & B1, MSX \citep{Price01} \\
4.356       &            90.04       &    0.01            & B2, MSX \citep{Price01}  \\
4.64         &           102.7        &   1.89            & M, \citet{Berrilli92} \\
8.276       &            141.2       &    0.01            & A, MSX \citep{Price01} \\
8.38         &           129.3        &   10.8            & N1, \citet{Berrilli92} \\
9.69         &           103.1        &   8.54            & N2, \citet{Berrilli92} \\
12.00       &            159.0       &    1.0             & IRAS \\
12.126     &             124.7      &     0.01           & C, MSX \citep{Price01} \\
12.89        &           104.2       &    9.88            & N3, \citet{Berrilli92} \\
14.649     &             104.8      &     0.01            & D, MSX \citep{Price01}\\
20            &          92.5        &    8.52            & Q, \citet{Simon74} \\
21.336     &             114.2      &     0.01            & E, MSX \citep{Price01} \\
25.00     &             224.0      &     1.0             & IRAS \\
350.0       &            4.18          &  0.35            & \citet{Mannings94} \\
450.0       &            2.16         &   0.11            & weighted average of \citet{Mannings94} and \citet{Sandell11} values\\
600.0        &           0.971        &   0.108          & \citet{Mannings94}\\
750.0        &           0.843         &  0.058           & \citet{Mannings94}\\ 
800.0        &           0.699         &  0.013           & \citet{Mannings94}\\ 
850.0         &           0.642          &   0.018         & weighted average of \citet{Mannings94} and \citet{Sandell11} values \\
1100.0        &          0.452         &  0.014           & \citet{Mannings94}\\
1300.0        &          0.1858       &   0.0049        & weighted average of \citet{Mannings94}, \citet{Manoj07}, \citet{Henning98}, and \citet{Alonso-Albi09}\\
1378.3        &          0.2887       &  0.02887       & \citet{Stapper22} \\
2600.0        &          0.149        &   0.005           & \citet{Alonso-Albi09} \\
6917.0        &          0.029        &   0.0010         & \citet{Alonso-Albi09} \\
13350.0      &           0.026        &   0.0002        & \citet{Alonso-Albi09} \\
36000.0      &           0.00939     &    0.00013     & weighted average of \citet{Skinner93} values \\
60000.0     &            0.01334      &   0.00036      & \citet{Skinner93} \\
\enddata

\tablenotetext{a}{adopted 0.02 as the overall uncertainty on all Pan-STARRS magnitudes, from \citet{Tonry12}}
\tablenotetext{b}{DENIS wavelengths from \citet{Fouque00}}
\tablenotetext{c}{2MASS values have been derived after applying corrections given by \citet{Cohen03}}
\tablenotetext{d}{MSX wavelengths from \citet{Cohen01}}

\label{litfluxes}
\end{deluxetable*}

\begin{figure}
 \hspace{1.75in}
\includegraphics[width=6.9cm,angle=-90]{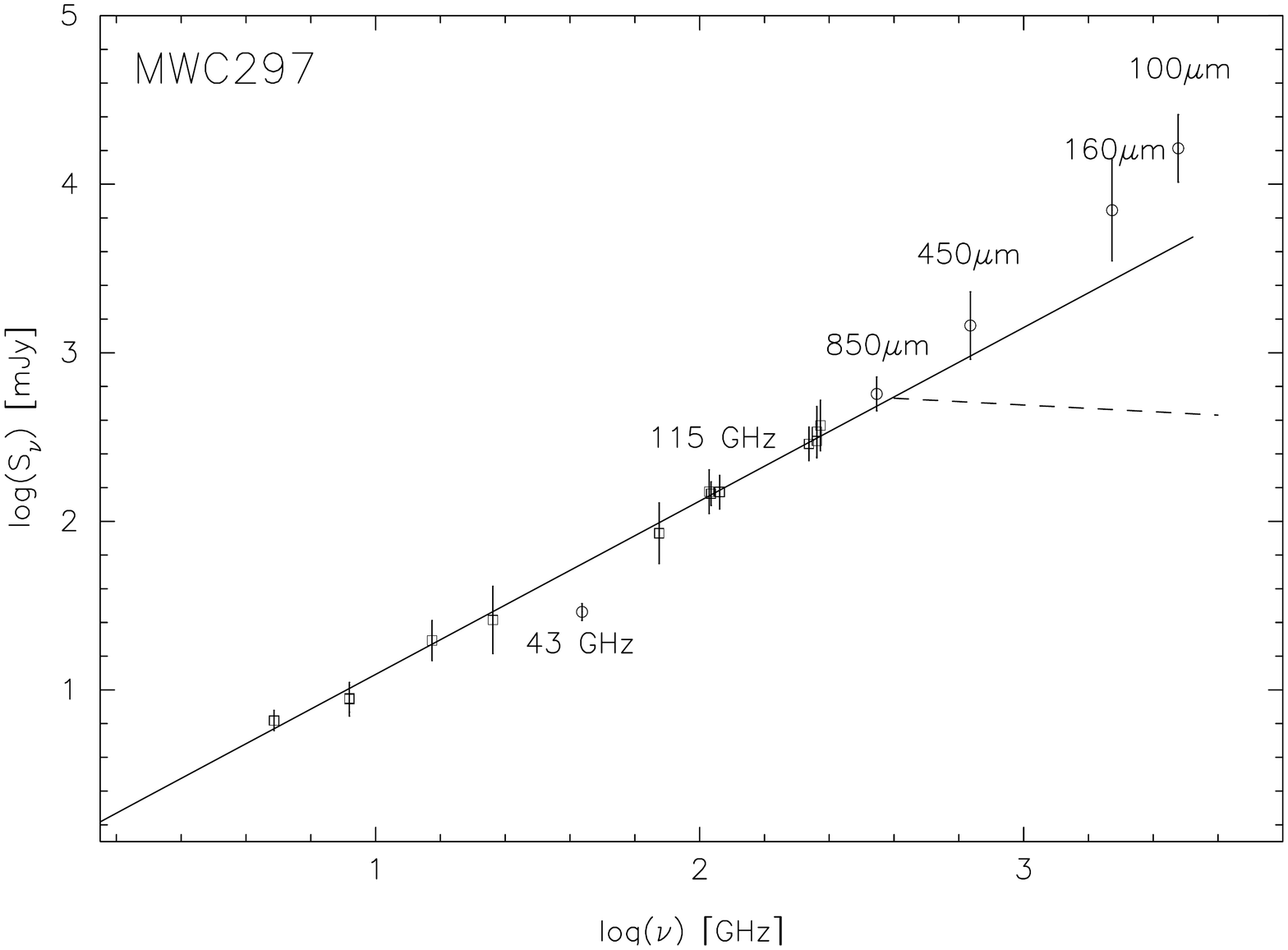}
\figcaption[]{
\label{fig-radiosed}
Least squares fit to the radio and 3 mm flux densities of MWC\,297, which yields a 
spectral index  $\alpha \sim$ 1.03. Data points plotted with circles were
omitted from the fit. The 850 $\mu$m and 450 $\mu$m flux densities are from \citet{Sandell11},
while the 160 $\mu$m and 100 $\mu$m PACS data are from this paper. The dot-dash line
shows an extension of the fit to shorter wavelengths.
There is a clear excess of flux (above the extension to the fit) at 100 $\mu$m and 160 $\mu$m, continuing 
down to about 450 $\mu$m. The turnover frequency appears to be between 850 and 450 $\mu$m. The dashed line shows
the case when the wind becomes optically thin at 400 GHz (750 $\mu$m). Values above the
dashed line are due to emission from warm dust.
}
\end{figure}

Both \citet{Manoj07} and \citet{Alonso-Albi09} found that the SED of MWC\,297 was rather
flat at mm-wavelengths, which they attributed to a population of large grains.
However, as pointed out by \citet{Sandell11} these authors severely underestimated
the free-free emission, which completely dominates the flux densities at millimeter 
wavelengths. Using all reliable VLA and 3 mm data, \citet{Sandell11} found that MWC\,297 has a spectral index  $\alpha \sim 1.03$.
We show a fit to the VLA and BIMA data described in this paper in Figure~\ref{fig-radiosed}, which is very similar to that
shown by \citet{Sandell11}. The similar fit is also shown by \citet{Rumble15},
who added their SCUBA-2 submillimeter flux densities to their plot. Figure~\ref{fig-radiosed} also demonstrates that there
is clearly excess flux due to warm dust at 100 $\mu$m and 160 $\mu$m. There
appears to be an excess at 450 $\mu$m as well, suggesting
that the wind becomes optically thin somewhere between 850 and 450 $\mu$m. 
With an estimate of the spectral index
of the ionized jet in hand, we can estimate the mass loss rate of ionized gas using the ionized jet model developed by \citet{Reynolds86}.
(See also eq. 3 in \citealp{Beltran01}.) The jet model requires estimates of 
the electron temperature, the velocity of the stellar wind, the jet opening angle, and inclination and the turnover frequency where
the jet becomes optically thin. We assumed T$_e = 10^4$  K, consistent with our adopted value for the SpeX analysis, and a 
wind velocity of $50$ \kms, as found from our best fit model of the SpeX spectrum.  
The jet inclination is 125\degr, as determined from our geometrical modeling of the outflow and the assumption that the disk where the jet originates is perpendicular to the outflow
(i.e., the disk inclination is the same as that of the outflow structure).
Based on the high angular resolution 6 cm MERLIN image we estimate opening angle of the jet where it is 
launched to be $\sim 32$\degr. If we assume a turn over frequency (the frequency at which the free-free emission becomes optically thin) 
of $\sim 400$ GHz \ref{fig-radiosed},
the jet model gives a mass loss rate of the ionized gas, 
$\dot{M} = 3.2 \times 10^{-7} M_\odot~\mathrm{yr}^{-1}$, which is within a  factor of two of the value derived from modeling the 
SpeX spectrum. The radius at which the ionized flow in the jet originates 
is found to be $\sim 19$ au. If a wind velocity of 75 \kms is adopted instead, then the predicted mass loss rate is  $\dot{M} = 4.8 \times 10^{-7} M_\odot~\mathrm{yr}^{-1}$,
in reasonable agreement with the value derived from our SpeX analysis. 

\subsection{The Northeastern Core} 

The strongest source detected at sub-mm and far
infrared wavelengths is a relatively compact core, $\sim$ 20\arcsec ~ to the
Northeast of MWC\,297, which we refer to as the NE Core, see
Table~\ref{tbl-PACS} and Figure~\ref{fig-PACS}. \citet{Rumble15} labeled it
SMM\,4\footnote{The Declination is incorrect for SMM\,4 in \citet{Rumble15}} and
it is core \#68 in \citet{Konyves15}. \citet{Rumble15} derive a temperature of
46 $\pm$ 2 K and a mass of 2.54 $\pm$ 0.14 \Msun\ (corrected to 418 pc), and
argue that it must be heated externally by MWC\,297. \citet{Konyves15} derive a
much lower temperature, 23.3 $\pm$ 0.2 K and a mass of 1.6 $\pm$ 0.1 \Msun, and
classify the core as being protostellar, even though there is no embedded source
in the core. We derive a smaller core size ($D \sim$ 13 -- 14\arcsec\} than both
\citet{Konyves15} and \citet{Rumble15} by careful background subtraction and
hence we derive lower flux densities. A gray body least squares fit gives a
temperature of 32 $\pm$ 1 K and a mass of 0.3 $\pm$ 0.05 \Msun. The
temperature we derive is in reasonable agreement with \citet{Konyves15},
considering that their automatic fitting routine almost certainly includes
emission from the surrounding cloud.

 \section{Discussion}
 
 \subsection{The rotation velocity and nature of the hot star in MWC297}

If the spectral type from \citet{Drew97} and the reddening estimate derived from our spectra and wind models are correct, then the SED we have generated
shows that the continuum flux throughout the infrared and even down into the optical wavelength range is predominantly due to dust and bound-free and possibly 
free-free continuum emission (Fig. \ref{fig-sed_full}). 
If the reddening is
$A_V \sim 7$ mag, somewhat lower than what we have derived ($A_V = 8.1$ mag) but still within our uncertainties, then a Kurucz model 
for a B1.5 V star would match the optical fluxes. However, dust, bound-free, and free-free continuum emission, perhaps due to shocks as a result of the accretion process, 
generate excess flux at wavelengths at least as short as $\sim 0.8 \mu$m and longwards. This is confirmed by our SpeX spectrum, which exhibits strong emission lines and continuum 
emission, with continuum discontinuities in emission for all three H series we detected, and no discernible sign of stellar absorption lines. It is also
consistent with the fractional dust contribution to the $H$ band flux of more than 90\% derived by \citet{Lazareff17}, although it is not clear that the dominant source 
of the excess emission is dust rather than bound-free or free-free emission \citep[see][]{Kluska20}.
This implies that attempts to derive the stellar parameters by fitting reddened stellar models directly to observed continuum 
fluxes beyond $\sim 0.8 \mu$m, and probably at shorter wavelengths as well, without attempting to account for the excess flux, may lead to incorrect results. This may 
explain why some values for the stellar luminosity, radius, and mass for MWC 297 given in the literature are so much larger than that 
expected for a B 1.5V star \citep[see e.\ g.][]{Vioque18,Ubeira20}. Adopting the smaller reddening value leads to a substantially smaller estimate of the
total luminosity and therefore a larger discrepancy between the estimated value and that expected for a B1.5V star. Adopting a stellar radius larger than $10 R_{\odot}$ to match the 
optical SED is inconsistent with both the parameters for a B1.5V star and the flux levels at the blue end of our SpeX spectrum.

Using our value of the inclination of the conical dust structure, assuming that the inclination of the central
star is the same as that of the outflow, and adopting the stellar rotation rate of $v \sin i = 350$ \kms ~found by \citet{Drew97} (under the assumption that the broad features detected
in the optical spectra are in fact rotationally broadened stellar photospheric lines), we find that the true rotational velocity
of the star is very high ($\sim 430$ km s$^{-1}$) and only slightly below the critical
velocity for a 10 $M_{\odot}$ star with $R \sim 6 R_{\odot}$ ($v_{crit} = (2GM_{\ast}/3R_{\ast})^{0.5} \sim 460$ km s$^{-1}$; \citealt{MaederMeynet2000}); the ratio
of the rotational velocity to the critical velocity is $\sim 0.9$. Furthermore, this value of the true rotational velocity would be well in excess of the critical velocity for a 10 $M_{\odot}$ star 
with any stellar radius $\geq 7 R_{\odot}$.
Such a high rotation rate would naturally result in an enhancement of the mass loss
at the equator. (See Fig 2 in \citealt{MaederMeynet10} for a depiction of the mass loss distribution for a star with properties very similar to those of MWC 297.) 
Therefore, even though MWC 297 appears to be still accreting, it could well be, simultaneously, in the throes of disruption due to its extreme rotational velocity. 
\citet{Acke08} also suggested that the mass loss in MWC 297 may be due to its high rotational velocity and speculated that it may eventually evolve into a classical Be star.

In this regard it is interesting to note the striking differences between the optical spectrum of MWC 297 obtained by \citet{Drew97} and \citet{Fairlamb15}, obtained in 1994 and 2009, respectively, and that of \citet{Andrillat98} obtained in 1997. The optical spectra obtained by \citet{Drew97} and \citet{Fairlamb15} are very similar, exhibiting strong Balmer emission lines sitting within deep absorption features. Photospheric absorption features
from C {\small III}, N {\small III}, and O {\small II}, which \citet{Drew97} used to determine both the spectral type and the rotational velocity, are present in the \citet{Fairlamb15}  spectra.
On the other hand, no photospheric H absorption features can
be discerned in the optical spectrum shown by \citet{Andrillat98}, which exhibits a multitude of weak emission lines in the $0.44-0.48 \mu$m wavelength region, none of which can be seen
in the spectrum presented by  \citet{Drew97}. The equivalent widths of the H $\alpha$ and H $\beta$ lines in the spectra from \citet{Drew97} are a factor of $\sim 2$ larger than
those reported by \citet{Andrillat98}. Similarly, the equivalent widths of the H  and O {\small I} lines given by \citet{Fairlamb17} are substantially and systematically larger than those 
listed by \citet{Andrillat98}. However, the equivalent width of the H $\alpha$ line reported by \citet{Acke05} in data obtained in 2002 was very similar to that measured by \citet{Andrillat98}.
We also note that \citet{Acke05} reported a change in the equivalent width of the [O {\small I}] 0.63 $\mu$m line of 25\% over a 3 month time period.


We have also retrieved the flux-calibrated X-shooter NIR spectrum obtained by \citet{Fairlamb15} from the archives and compared it with our SpeX spectrum. The X-shooter spectrum was obtained two years after our SpeX observations. This comparison reveals differences in the strength of both the continuum and the emission lines. The 
continuum in the $J$ and $H$ bands is a factor of $\sim 2$ stronger in the SpeX data compared to that in the X-shooter spectrum. Similarly, the Pa $\beta$ line flux is nearly a factor of 2 larger in our SpeX spectrum, while the Br $\gamma$ line, which is quite strong in our SpeX data, is nearly indistinguishable above the noise in the spectrum obtained by \citet{Fairlamb15}. \citet{Fairlamb17} give only an upper limit for flux of the Br $\gamma$ line that is a factor of 12 lower than the flux we directly measure in our SpeX spectrum.
The O {\small I} 0.8446 $\mu$m line is somewhat stronger in the X-shooter spectrum than in our SpeX data. \citet{Fairlamb17} also report fluxes for the Ca {\small II} triplet at 
0.85 $\mu$m that are significantly larger than the fluxes we measure in the nearby upper Paschen series; we see no evidence of Ca {\small II} emission in our SpeX spectrum. (\citealt{Andrillat98} also did not detect the Ca {\small II} lines.)
Given the differences in H line strengths, it is not surprising that the mass
accretion rate we derive from the strongest emission line strengths in the SpeX spectrum is considerably larger than that reported by \citet{Fairlamb17}, even after accounting
for the different distances used in the calculation. We also note that
although the He {\small I} $1.083 \mu$m 
line in our spectrum exhibits a P Cygni profile (with two blue absorption features), indicating formation in an outflowing wind, we see no sign of absorption at He {\small I} 2.058 $\mu$m, as reported by \citet{Murdoch94} (although the absorption could possibly be masked by residuals from the telluric correction in our spectrum).

The optical photometry of MWC 297, as well as the fluxes of the strongest H emission lines reported
in the literature, combined with the results of this comparison indicate considerable, rapid variability in both the continuum and the emission line strengths
of the source. It is possible that this irregular 'flickering' of the source strength in MWC 297 is due to the unstable nature of the central star as a result of
its high rotational velocity, extremely close to the break-up velocity. Examination of the {\it Herschel} PACS image \ref{fig-PACS_3color}, as well as the WISE, Spitzer, and 
2MASS images of MWC 297 reveal numerous arcs of emission at a range of radii up to $\sim 1.5$ pc from the central star in MWC 297 \citep[see e.\ g.][]{Wang07}. 
The appearance of these 
structures, combined with the variability of the accretion and mass-loss rates (as evidenced by the variable H emission line strengths), suggests that MWC 297 has 
experienced multiple mass ejection events since its formation. It is tempting to speculate that these ejection events are due to the high rotational velocity of
the star, perhaps combined with the orbital effects of the close companions discovered by \citet{Ubeira20} and \citet{Sallum21}, which may be modulating the accretion rate from 
the central disk.

\subsection{The accretion disk in MWC 297}

Observations of double-peaked CO lines in NIR spectra \citep{Banzatti22,Sandell22} clearly indicate that an accretion disk is present in MWC297. However,
the stellar rotational velocity, corrected for the inclination, indicates that the disk co-rotation radius, $R_{cor} = (GM_{\ast}R_{\ast}/v_{\ast}^2)^{1/3}$, 
must be $\sim 7 R_{\odot}$. Since the disk truncation radius must be smaller than the co-rotation radius, the accretion disk must extend down to nearly the stellar 
surface in MWC 297.\footnote{Note that requiring that $R_{cor} \ge R_{\ast}$ places a firm upper limit on the stellar radius of $R_{\ast} \le 10 R_{\odot}$ for
a $10 M_{\odot}$ star with the estimated rotation velocity.} \citet{Acke08} and \citet{Kluska20} reached a similar conclusion from modeling their interferometric data on MWC297,
which failed to reveal an inner gap or cavity between the disk and the star. This result strengthens the claim that the model of magnetospheric accretion, widely accepted 
for lower mass Classical T Tauri stars, is probably not applicable to Herbig Be stars \citep[see e.g.][]{Mottram07,Mendigutia11,Wichittanakom20}. 
The boundary layer theory of accretion, proposed by \citet{Lynden74}, accompanied by a radiation-driven wind from either the star itself or the disk, 
as proposed by \citet{Drew98}, appears to be the more appropriate model for high luminosity objects like MWC 297.

 It is unclear why the orientation of the system, derived directly from our MIR images, is so
 different from the values derived from interferometric analyses. While it is beyond the scope of this paper to investigate the reasons for this discrepancy,
 we can speculate about possible causes. It is possible that
 the inner source has rotated and the larger scale orientation that we observe in our FORCAST images does not reflect
 that of the inner disk (i.e., the inner disk is tilted.) However, given the young age of the system ($\sim 0.1$ Myr) and the magnitude of the inclination discrepancy ($> 15$ \degr),
  it is not clear how such a large misalignment of the disk and the larger-scale structure would have occurred in such a short time. The examples of systems with misaligned inner 
  disks among HAeBe stars are generally fairly old (e.g., AB Aur and HD 100546, which have ages of several million years; \citealt{vandenAncker98, DeWarf03}).
 Furthermore, the radio data, which traces the outflow from its origin in the inner disk and therefore presumably does reflect the orientation of the inner disk, 
 seems to be well aligned with the structure we see in the FORCAST images. In addition, \citet{Sallum21} have derived disk inclinations similar to those
 we have found from our direct MIR observations from analysis of their interferometric data when they
 include an outflow along with an inclined disk in their models. They claim that the disk plus outflow model provides the best representation of the observed brightness distribution
 in their data. Their modeling also yields a position angle of the outflow that is similar to ours.
 
 A possibly more likely explanation for the discrepancy in the inclination values is that the 
 interferometric analyses are in error simply because the assumptions involved in the 
 modeling the data are incorrect or incomplete. \citet{Malbet07} modeled AMBER/VLTI observations of MWC 297 in the Br $\gamma$ line using a combination of an optically
 thick gas disk and an optically thin wind. However, our models suggest that the latter assumption is invalid in MWC 297, for which all of the NIR lines appear to be
 optically thick, and Br $\gamma$ extremely so.  Interferometric studies often find that the Br $\gamma$ emitting region in MWC 297 is more extended than the 
 region that generates the continuum emission \citet[e.g.][]{Malbet07,Kraus08,Hone17}. Most models used to interpret these results assume that the continuum 
 emission is due to dust in the accretion disk close to the star \citep[e.g.][]{Millan-Gabet01,Eisner04,Acke08,Kraus08}. These models usually encounter difficulties
 as they yield locations for the emission regions that are so close to the star that the temperature of the dust which is supposed to be responsible for the emission 
 is beyond the sublimation limit. Our SpeX data, in which strong 
 continuum jumps due to bound-free emission can be clearly seen, suggest that the assumption that emission from dust is the predominant component of the NIR 
 flux may be incorrect. Instead, an optically thick gaseous region close to the star may be responsible for this flux via bound-free or free-free emission \citep[see also][]{Acke08,Kluska20}. 
 \citet{Muzerolle04} has shown that emission from an optically thick gaseous inner
 disk region may be a substantial contributor to the NIR flux in sources with accretion rates as high as that we estimate for MWC 297.
 Our results, combined with those of \citet{Sallum21}, suggest that interpretation of interferometric data needs
 to be approached with some caution, as models which do not include important physical aspects of the system 
 may lead to erroneous results.

 \section{Summary and Conclusions}
 
 Our SOFIA/FORCAST images at 20-40 $\mu$m provide a unique and direct view of the dust
surrounding the massive Herbig Be star MWC 297. We interpret the extended MIR
emission seen in the FORCAST images as arising from warm ($T_{dust} \sim 67$ K),
optically thin dust entrained at the edges of the bi-polar outflow that is seen in our VLA and BIMA
radio data. The structure seen in the MIR images reflects the conical shape expected from such a
scenario. The southern arc is in the foreground, tipped toward the observer,
while the northern annulus is tipped away. We constructed a simple geometrical model of the system
and generated simulated MIR images to constrain the physical parameters. 
We find the inclination of the system to be $i \sim 55$ \degr, which directly contradicts the low
inclination values derived from the analysis of the previous interferometric
observations. Nevertheless, our derived inclination value agrees with the approximately N-S
extension of the ionized flow seen in the radio maps of \citet{Sandell11} as well as
the recent results of the interferometric images and analysis by \citet{Sallum21}.
The inclination value implies that the central star in MWC 297 must be rotating close to its break-up velocity. Furthermore, 
the inclination allows us to estimate the truncation radius of the accretion disk, which must be close to the stellar surface.
This result provides further evidence that the magnetospheric accretion model is not applicable to high mass Herbig Be stars.
 
Our NIR spectrum of MWC 297 obtained with SpeX reveals a host of strong emission lines, most from the Paschen, Brackett, and
Pfund series of H, and continuum emission that rises to longer wavelengths. Continuum jumps from the H series are seen in
emission. Although the He {\small I} $1.083\mu$m line exhibits a P Cygni profile, no other absorption features are seen in this spectrum.
We modeled the H line emission with a wind model and find that all of the lines are optically thick. The analysis yields a mass-loss
rate for the system of $\sim 6 \times 10^{-7} M_{\odot} {\rm yr}^{-1}$ and an extinction of $A_V = 8.1$ mag. The ionization rate needed
to generate these lines is far above that expected from a B1.5V star atmosphere, and therefore most of the ionizing flux must arise
from the accretion process.

We compiled data from the literature and combined it with our FORCAST, VLA, and BIMA measurements and archival {\it Herschel} data to
generate the SED of MWC 297 from $0.35~ \mu$m to $6$ cm. Integrating under the SED, we calculate a total luminosity of 
MWC 297 of $\sim 6900 L_{\odot}$, somewhat less than that expected for a B1.5V star, although we are certainly missing flux from unobserved
wavelength regions. After accounting for flux between 0.25 and 0.35 $\mu$m, we estimate a bolometric luminosity of MWC 297 of $\sim 7900 L_{\odot}$,
with an upper limit of $\sim 10200 ~L_\odot$
The SED analysis reinforces the suggestion of \citet{Acke08} that the extinction we measure from the H lines with the wind model
is due to foreground dust, distant from the star, and not the dust we detect in the immediate vicinity.

Finally, we speculate that the high rotational velocity ($\sim 430$ km s$^{-1}$, or $\sim 90$\% of the break-up velocity) of the central star in MWC 297, 
perhaps combined with the orbital effects of the two possible nearby low-mass companions recently detected via interferometry, may be responsible for 
the rapid fluctuations in the mass accretion and mass-loss rates from the central disk. These fluctuations would then produce the variations observed in the   
continuum and emission line strengths.

%
%

\section*{acknowledgements} 	
This work is based in part on observations made with the NASA/DLR Stratospheric Observatory for Infrared Astronomy (SOFIA). 
SOFIA is jointly operated by the Universities Space Research Association, Inc. (USRA), under NASA contract NNA17BF53C, and 
the Deutsches SOFIA Institut (DSI) under DLR contract 50 OK 2002 to the University of Stuttgart. Financial support for this work 
was provided by NASA through award \#02-0016 issued by USRA. 
The National Radio Astronomy Observatory is a facility of the National Science Foundation operated
under cooperative agreement by Associated Universities, Inc. The BIMA array
was operated by the Universities of California (Berkeley), Illinois, and
Maryland with support from the National Science Foundation. This research
has made use of data from the Herschel Gould Belt survey (HGBS) project. 
The HGBS is a Herschel Key Programme jointly carried out by SPIRE Specialist Astronomy Group 3 (SAG 3),
scientists of several institutes in the PACS Consortium (CEA Saclay,
INAF-IFSI Rome and INAF-Arcetri, KU Leuven, MPIA Heidelberg), and
scientists of the Herschel Science Center (HSC). We thank the referee, whose comments helped us to
clarify certain points in the text and improve the manuscript.
Finally, we would like to thank John Rayner for acquiring the SpeX spectrum for us and Richard Plambeck for acquiring and analyzing  the BIMA data 
used in this paper and for a careful reading of the final draft.
 


\startlongtable
\begin{deluxetable*}{cccl}
\tabletypesize{\scriptsize}
\tablecolumns{4}
\tablenum{1}
\tablewidth{0pt} 
\tablecaption{MWC\,297 line fluxes measured with IRTF/SpeX \label{tbl-5}} 
\tablehead{
\colhead{Obs. Wavelength}   &  \colhead{Obs. Flux} & \colhead{Uncertainty} & \colhead{ID} \\
\colhead{($\mu$m)} &  \multicolumn{2}{c}{($10^{-16}$ W m$^{-2}$)} & 
}
\startdata
0.832663 &    2.53 &    0.66 & Pa 25 \\
0.833688 &    5.48 &    0.91 & Pa 24 \\
0.834848 &    5.97 &    0.62 & Pa 23\\
0.836198 &    8.37 &    0.48 & Pa 22 \\
0.837740 &   10.30 &    0.47 & Pa 21 \\
0.839541 &   11.87 &    0.45 & Pa 20 \\
0.841625 &   15.83 &    0.48 & Pa 19 \\
0.844929 &  261.57 &    0.64 &  Pa 18 + O I\\
0.847015 &   21.92 &    0.46 & Pa 17  \\
0.850542 &   24.34 &    0.43 & Pa 16  \\
0.854833 &   27.52 &    0.41 & Pa 15 \\
0.860108 &   34.89 &    0.41 &  Pa 14\\
0.863217 &    6.34 &    0.33 &  N I\\
0.866803 &   37.41 &    0.38 & Pa 13 \\
0.868566 &   18.80 &    0.41 &  N I\\
0.875349 &   42.55 &    0.36 & Pa 12 \\
0.886589 &   52.12 &    0.35 & Pa 11 \\
0.901800 &   61.24 &    0.39 &  Pa 10\\
0.906406 &    5.18 &    0.31 &  \\
0.909760 &    6.03 &    0.36 &  C I\\
0.923224 &   86.59 &    0.44 &  Pa 9\\
0.924724 &   10.70 &    0.39 &  Mg II\\
0.926710 &   10.85 &    0.44 &  O I\\
0.954920 &  131.62 &    0.53 & Pa $\epsilon$ \\
1.000116 &   13.00 &    0.22 &  \\
1.005296 &  233.13 &    0.45 & Pa $\delta$ \\
1.011420 &   14.02 &    0.26 &  \\
1.017763 &    1.80 &    0.21 &  \\
1.029057 &    2.14 &    0.21 &  [S II] \\
1.032398 &    2.68 &    0.16 & [S II] \\
1.043853 &    1.19 &    0.24 &  \\
1.046021 &    2.37 &    0.21 &  \\
1.050532 &   10.88 &    0.21 &  \\
1.053146 &   19.74 &    0.63 &  \\
1.072321 &   54.94 &    0.81 &  \\
1.083737 &   23.18 &    0.25 &  He I\\
1.086643 &   11.22 &    0.23 &  \\
1.091859 &   12.89 &    0.27 &  \\
1.094196 &  528.52 &    0.90 &  Pa $\gamma$ \\
1.112957 &   10.71 &    0.33 &  \\
1.116940 &   12.51 &    0.83 &  \\
1.129121 &  446.46 &    0.82 &  O I\\
1.175703 &   27.07 &    0.23 & C I \\
1.189867 &    4.80 &    0.18 &  \\
1.200254 &    1.71 &    0.18 &  \\
1.213768 &    2.38 &    0.21 &  \\
1.219186 &    4.12 &    0.20 &  \\
1.220817 &    2.85 &    0.21 &  \\
1.233349 &    4.31 &    0.20 &  \\
1.238663 &    6.81 &    0.20 &  \\
1.240800 &    2.10 &    0.22 &  \\
1.246987 &   31.59 &    0.29 &  \\
1.257211 &   12.21 &    0.42 &  \\
1.261574 &    7.45 &    0.38 &  \\
1.282257 & 1588.64 &    1.78 &  Pa $\beta$\\
1.316919 &   89.94 &    0.36 &  O I\\
1.343495 &   12.81 &    0.37 &  \\
1.454872 &   16.53 &    0.37 &  \\
1.482794 &   23.14 &    1.66 &  \\
1.485319 &    5.42 &    0.31 &  Br 30\\
1.487256 &   10.83 &    0.41 &  Br 29\\
1.489312 &   12.97 &    0.32 &  Br 28\\
1.491721 &   14.37 &    0.31 &  Br 27\\
1.494332 &   18.95 &    0.32 &  Br 26\\
1.497276 &   25.21 &    0.33 &  Br 25\\
1.500623 &   27.39 &    0.33 &  Br 24\\
1.504437 &   36.56 &    0.37 &  Br 23\\
1.508821 &   38.67 &    0.36 &  Br 22\\
1.513864 &   47.09 &    0.36 &  Br 21\\
1.519726 &   54.77 &    0.36 &  Br 20\\
1.526598 &   58.65 &    0.36 &  Br 19\\
1.534724 &   66.29 &    0.38 &  Br 18\\
1.544442 &   78.16 &    0.39 &  Br 17\\
1.556195 &   88.31 &    0.41 &  Br 16 \\
1.558674 &   11.62 &    0.33 &  \\
1.570623 &   98.93 &    0.44 &  Br 15\\
1.588622 &  123.30 &    0.49 &  Br 14\\
1.611500 &  150.41 &    0.56 & Br 13 \\
1.641303 &  157.17 &    0.61 & Br 12 \\
1.681246 &  164.14 &    0.71 &  Br 11\\
1.682854 &   81.20 &    1.90 &  \\
1.703959 &   11.72 &    1.35 &  \\
1.736831 &  187.78 &    0.97 &  Br 10\\
1.945248 &  380.34 &    1.78 &  Br 8\\
2.059333 &   22.94 &    1.29 &  He I\\
2.166293 &  512.79 &    1.42 &  Br $\gamma$\\
2.329991 &    8.05 &    1.05 &  Pf 34\\
2.333117 &   10.30 &    1.28 &  Pf 33\\
2.336679 &   13.03 &    1.44 &  Pf 32 \\
2.340345 &   16.19 &    1.46 &  Pf 31\\
2.344606 &   10.74 &    1.03 &  Pf 30\\
2.349666 &   17.21 &    1.46 &  Pf 29\\
2.354513 &   25.41 &    1.52 &  Pf 28\\
2.360591 &   20.14 &    1.22 & Pf 27 \\
2.367197 &   28.05 &    1.56 &  Pf 26\\
2.374560 &   27.92 &    1.64 &  Pf 25\\
2.383186 &   28.52 &    1.41 &  Pf 24\\
2.392506 &   30.58 &    1.32 &  Pf 23\\
2.403738 &   31.52 &    1.37 & Pf 22 \\
2.416510 &   42.94 &    1.69 & Pf 21 \\
\enddata
\label{SpeXfluxes}
\end{deluxetable*}




\end{document}